\documentclass[a4paper]{amsart}
\usepackage{amsthm}
\usepackage{amsmath}
\usepackage[hmarginratio={1:1},vmarginratio={1:1}]{geometry}
\usepackage{amsfonts}
\usepackage{graphicx}
\usepackage{array}
\usepackage{amssymb}
\usepackage{upgreek}
\usepackage{mathrsfs}
\usepackage{float}

\usepackage{pgf}
\usepackage{tikz}
\usetikzlibrary{arrows}
\usetikzlibrary{cd}
\usetikzlibrary{matrix}
\usetikzlibrary{positioning}
\usepackage{color}
\usepackage{empheq}

\usepackage{amsaddr}

\newtheorem{theorem}{Theorem}
\newtheorem{corollary}[theorem]{Corollary}
\newtheorem{proposition}[theorem]{Proposition}

\numberwithin{equation}{section}
\numberwithin{theorem}{section}

\begin{document}

%\markboth{A. Bautista}{Sky conformal invariant}

\title[Opinion dynamics with manipulative agents]{\textbf{Opinion dynamics in bounded confidence models with manipulative agents: Moving the Overton window.}}

%%% Para \documentclass{amsart}
%%%%%%%%%%%%%%%%%%%%%%%%%%%%%%%%
\author{Alfredo Bautista}
\address{\small Dpto. de An\'alisis Econ\'omico: Econom\'{\i}a Cuantitativa, Universidad Aut\'onoma de Madrid \\ C/ Francisco Tom\'as y Valiente 5, 28049 Madrid, Spain. }
\email{alfredo.bautista@uam.es}
%\date{}
%\subjclass{Primary 05C38, 15A15; Secondary 05A15, 15A18, ???}

\begin{abstract}
This paper focuses on the opinion dynamics under the influence of manipulative agents. This type of agents is characterized by the fact that their opinions follow a trajectory that does not respond to the dynamics of the model, although it does influence the rest of the normal agents. Simulation has been implemented to study how one manipulative group modifies the natural dynamics of some opinion models of bounded confidence.
It is studied what strategies based on the number of manipulative agents and their common opinion trajectory can be carried out by a manipulative group to influence normal agents and attract them to their opinions. In certain weighted models, some effects are observed in which normal agents move in the opposite direction to the manipulator group. Moreover, the conditions which ensure the influence of a manipulative group on a group of normal agents over time are also established for the Hegselmann-Krause model.
\end{abstract}

\maketitle

\tableofcontents

\section{Introduction}

The opinion dynamics is a research topic that has been quite active for several decades from various points of view: economical \cite{Fo74,Tr24}, political \cite{Be05,Ro22}, mathematical \cite{Lo07,Bl10}, sociological \cite{Wa99,Lo18} among others \cite{Ja05,St18}. 
Its study focuses on the phenomena that appear as a result of the interaction between the opinions of different agents. 
To this end, different types of mathematical models \cite{DG72} have been developed which, by their nature, can only represent very specific aspects of the mechanisms involved in the formation of opinion. 
Such models usually implement rules to describe the interactions between the agents involved, which may correspond to individuals, advertising messages, bots on social networks, or groups of any of these. 
In addition, the opinions of the agents can obey the general rules of each model or particular rules that only affect them, even if the rest of the agents interact with them following the general rules. 

This can be seen most clearly when you consider that such special agents are charismatic leaders, advertising agents, or even fake news on social media who try to convince normal agents with their opinions or messages that are usually constant over time. 
There may also be situations where an agent wants to change their mind drastically without losing influence over normal agents who already rely on it. 
To do this, it will gradually modify its opinion until it arrives at the new and intended one.

For example, politicians who want to carry out a totally different policy than the one they have pursued so far, risk losing their voters. To avoid this, they can change their message over time, convincing their voters of the need for small changes, without ever discovering their ultimate intentions. After that period of time, the sum of those small changes is the big change of opinion desired to be able to propose their new policy.
This example also illustrates a way of making policies acceptable that were once radical or unacceptable, thus recalling the Overton Window theory. This theory posits that policies acceptable (debatable) by society are within a window (range) and in order for society to accept other radical policies it is first necessary to widen or displace this window until society deems them acceptable.
For all these reasons, it is interesting to study how the opinion dynamics evolve when in a society there are this type of special agents, which we will call \emph{manipulators}, as well as strategies that they can carry out to achieve their objectives.

In this paper we will study the influence of groups of manipulative agents on discrete time bounded confidence models such as the Deffuant--Weisbuch \cite{De00} and Hegselmann--Krause \cite{He02} models as well as the weighted bounded confidence models developed in \cite{To20,Mi21}. The models studied also have dynamics that lead to different types of final distributions: a single central cluster (consensus), multiple clusters distributed throughout the opinion space, two clusters at the extremes (polarization) or even oscillating opinions in a certain range of the opinion space. For this reason, there is no general definition of the Overton Window that can be suitable for all types of final distribution. 

The article is organized as follows. Section \ref{sec:Generalmodel} is devoted to introduce the motivations in which this paper is based and a general model with manipulative agents. 
The following sections, which are largely independent of each other, deal with the various models that we will study in the article.
In section \ref{sec:HKclassic}, we will describe the Hegselmann--Krause model with and without manipulative agents. The interaction between manipulative and normal agents is mathematically analysed and a simulation showing the ability of manipulators to modify normal opinions is presented. The study of this interaction is based on the number of manipulative agents and the speed at which they change their opinion. 
In a similar way, in section \ref{sec:DefWeiMod}, the Deffuant--Weisbuch model is described and the influence of manipulators is simulated under analogue parameters. 
Section \ref{sec:WHKmodel} describes three types of Weighted HK models: Attractive, Repulsive and Attractive--Repulsive. Again, the influence of one manipulative group is studied by simulation.
Finally, in section \ref{sec:conclusion}, the results obtained and the work that can be carried out in future articles are discussed.

\section{Bounded confidence models with manipulative agents}\label{sec:Generalmodel}

The aim of this section is to introduce the general opinion dynamics model with manipulative agents as a modification of already known bounded confidence agent--based models.

\subsection{Opinion dynamics without manipulative agents}\label{sec:BCmodels}

Let us consider a discrete agent space $P = \{p_i\}_{i=1}^{N}$. Each agent $p_i\in P$ is a subject having an opinion about some issue. The opinions can be parametrized by some real number $x\in \mathbb{R}$ (that can be thought as the amount of resources, capital, etc. that the agent is willing to invest or increase from a budget), in such a way that $x(p_i) = x_i$. 
We can consider that these values are all contained in an interval $[a,b]\subset \mathbb{R}$ called the \emph{opinion space}, which can be interpreted as the range of all possible opinions.
The opinion of each agent can vary along the discrete time $t\in \mathbb{N}\cup \{0\}$, so we can denote by $x_i(t)$ the opinion of the agent $p_i$ at time $t$ and $\mathbf{x}(t)=\left(x_1(t),\ldots ,x_N(t)\right)\in \mathbb{R}^{N}$.
The main feature of \emph{bounded confidence models}, or \emph{BC--models} for short (see \cite{He02}), is that each agent can only be influenced by agents with close opinion. 
Indeed, for the agent $p_i\in P$ and for some value $\varepsilon_i >0$, called \emph{confidence threshold}, we can denote the \emph{set of confidence of $p_i$} by $J_i(t)\in \mathbb{R}$ which can be defined by 
\begin{equation}\label{eq-set-of-confidence}
J_{i}(t)=\left\{ k\in\{1,\ldots,N\} : \vert x_i(t)-x_k(t) \vert \leq \varepsilon_i \right\}
\end{equation}
and it is made up of the indexes of agents who have their opinion at a distance no higher than the confidence threshold $\varepsilon_i$ of the agent $p_i$. The number of agents in $J_{i}(t)$ will be denoted by $\vert J_{i}(t) \vert$. Observe that, $\vert J_{i}(t) \vert > 0$ because $i\in J_{i}(t)$ for all $i=1,\ldots,N$. 
The dynamics of the opinion distribution $\{x_i(t)\}_{i=1}^{N}$ is described, for each $x_i$, by an iteration such that 
\begin{equation}\label{eq-opinion-dynamic-general}
x_i(t+1) = x_i(t) + F_i\left(\mathbf{x}(t),t\right) 
\end{equation}
where the functions $F_i$ depend only on the opinions $x_j(t)$ where $j\in \Omega_i(t)\subset J_i(t)$. The set $\Omega_i(t)$ is defined as the set of indexes of agents which influence the opinion of agent $p_i$. So, the opinion $x_i(t+1)$ is defined by the iteration (\ref{eq-opinion-dynamic-general}) only if $j\in\Omega_i(t)$, otherwise $x_i(t+1)=x_i(t)$. This implies that the sets $\Omega_i$ and the functions $F_i$ characterize the model defining the variation of the opinion of $p_i$ at the discrete time $t$.

In addition to the choice of the sets $\Omega_i$ and the functions $F_i$, there are many ways to modify these models to adapt them to different cases. For example, the number of modified opinions at every iteration $t$ leads to different models, so we will say that a model is \emph{simultaneous} if the iteration given in (\ref{eq-opinion-dynamic-general}) is applied to all agents in order to modify their opinions for each $t$. On the other hand, we will call a model \emph{alternating} if, for each $t$, iteration (\ref{eq-opinion-dynamic-general}) is applied to a single random agent.

\subsection{Types of agent}\label{sec:TypAge}

In \cite{He15,Mi21,Bo21}, several types of agents are defined for the BC--model depending on where their opinions are situated in the opinion space, whether or not their opinions evolve under the dynamics equation (\ref{eq-opinion-dynamic-general}) and their influence over the rest of agents. In this article, we will consider the following types of agents:
\begin{itemize}
\item A \emph{normal} agent is that whose opinion evolves as $x_i$ under the dynamics equation (\ref{eq-opinion-dynamic-general}). They establish the general type of agents and influence each other homogeneously. 
\item A \emph{stubborn} agent is that whose opinion do not change over time. The opinions of these agents influence others but are not influenced by the rest. So, if $p_j$ is a stubborn agent, then $x_j(t+1)=x_j(t)$ for all $t\in \mathbb{N}\cup \{0\}$
\end{itemize}

If there are $1<K\in\mathbb{N}$ stubborn agents with the same opinion in the BC--model, they can be considered as a group of agents whose influence on the neighbouring agents is equivalent to the one of an only agent with the corresponding weighted opinion. Such groups can be seen as an information campaign rather than a group of actors, so they influence society but are not counted as part of it. 

Let us introduce a new type of agents that will not follow equation (\ref{eq-opinion-dynamic-general}).
\begin{itemize}
\item A \emph{manipulative agent} or \emph{manipulator} is an agent $p_i$ whose opinion is not influenced by the rest of agents but follows a non--constant path $x_i(t) = f_i(t)\in [a,b]\in \mathbb{R}$, where $f_i$ is a function with values in the opinion space $[a,b]\in \mathbb{R}$ that will be defined prior to the implementation of the model.  
\end{itemize}

Manipulative agents can be thought as a generalization of stubborn agents allowing them to change their opinions without being influenced by the rest of society. This change of opinion may aim to drag the opinion of normal agents close to a limit cluster that, \emph{a priori}, was not expected to reach.

\subsection{The Overton window}\label{sec:Overtonw}

In the 1990s, J. P. Overton introduced a political theory, later called the Overton window, according to which government policies can be classified within a spectrum that describes their acceptability by society. It establishes a classification of opinions in which only a small range is considered acceptable. Outside that acceptable range, opinions can be classified as radical or even unthinkable. Government policies will always be in the range of acceptability, otherwise the government will lose voters or popularity. In this theory, it is the opinions within the Overton window that enable political action. Therefore, the shift and expansion or contraction of the Overton window (of acceptable opinions) can cause previously accepted opinions to become radical or unthinkable and \emph{vice versa}, affecting the policies that the government will be able to carry out.

In relation to opinion models, we can think of the Overton window of acceptable opinions as a interval $[c,d]$ contained in the opinion space $I=[a,b]$ of all possible opinions, that is $[c,d]\subset I$. This subinterval can be characterized by two numbers: a center value and a width. But notice that there is no canonical way to define the limits of acceptable opinions and it also depends on the existence of a number of agents who have the same or similar opinion.

The dynamics of the model, at every iteration, will change the initial opinion distribution and thus also the measurements that characterize some properties: mean value, the position of possible final clusters as well as their number of agents contained in them. These characteristics can be shifted in the opinion space, just as dispersion can increase or decrease, depending on the behaviour of the possible stubborn or manipulative agents. So, these changes in statistical values are reminiscent of the movement of the Overton window in the political theory. 

Now, some questions arise. We may wonder what strategy a group of manipulative agents can carry out to move the Overton window in the direction that is favourable to their interests. Is it possible to quantify the size of the group of manipulators so that they can have enough influence to drag the opinion of normal agents to their target?

In section \ref{sec:HKmodel}, after introducing the Hegselmann--Krause model with manipulative agents, we will illustrate how a group of manipulators is able to influence the opinion of society to shift the range of opinions towards one of the extremes of the opinion space, even to move many opinions to the most radical possible opinion (see figure \ref{figure-scBCOver}).

\subsection{A general model with manipulative agents}\label{sec:GenModMan}

According to the definition of manipulative agent in section \ref{sec:TypAge}, the opinion of the manipulators is not under the dynamics of the models, but follows a trajectory previously defined for each time $t$. Therefore, the manipulators can influence the normal agents, but the normal agents will not be able to influence the manipulators.

So, we will introduce manipulative agents in a general model. Let us fix the opinion space to $I=[-1,1]$. Consider an agent space $P = \{p_i\}_{i=1}^{N}$ with distribution of opinion at time $t$ given by $\{x_i(t)\}_{i=1}^{N}$ and $K\in\mathbb{N}$ manipulative agents with predefined opinions $\{f_j(t)\}_{j=1}^{K}$ for time $t$. Now, the extended set of confidence $\overline{J}_i(t)$ of the normal agent $p_i$ is defined by 
\begin{equation}\label{eq-J-manipulators}
\overline{J}_{i}(t)=\left\{ k\in\{1,\ldots,N+K\} : \vert x_i(t)-z_k(t) \vert \leq \varepsilon_i \right\}
\end{equation} 
where $z_k(t)\in \{x_i(t)\}_{i=1}^{N} \cup \{f_j(t)\}_{j=1}^{K}$ can be normal or manipulator and defined by
\begin{equation}\label{eq-zk-definition}
z_k(t)=\left\{ 
\begin{array}{lcl}
x_k(t) &  &  \text{ if } k\leq N  \\
f_{k-N}(t) &  &  \text{ if } k> N
\end{array}
\right.   .
\end{equation}
Thus, the subset $\Omega_i(t)\subset \overline{J}_i(t)$ of neighbouring agents influencing $i$--th normal agent can contain both manipulative and normal agents. 

We are interested in the study of the opinion dynamics of normal agents $\{x_i(t)\}_{i=1}^{N}\subset [-1,1]$ at time $t+1$, which will depend on its own opinion $x_i(t)$ at time $t$ and the opinions $z_j(t)$ of the neighbouring agents for $j\in \Omega_i(t)$, whether normal or manipulative. 

We will assume that the manipulators are in an organized group such that all their agents have always the same opinion and they change it at every iteration. With this, they achieve a greater influence on the normal opinions of their confidence interval because their common opinion has a weight of $K$ manipulative agents in the interval.

Throughout this article, in addition, we will study the opinion dynamics in simultaneous models because in this way we are avoiding the random component of choosing, for each iteration $t$, a normal agent on which to apply the model. Therefore, each iteration will be applied to all normal agents at the same time. 

In simultaneous implementations, the set $\Omega_i(t)$ can be interpreted as the subset of agents having an effective influence on the opinion $x_i(t)$ in the interval of time $[t,t+1]$, so some others agents in the confidence interval can interact with the agent $p_i$ but they do not have real influence on it.

Thus, we can study the influence of a group of manipulators on the opinion of normal agents according to the dynamics of diverse variants of the classic models such as Hegselmann--Krause \cite{He02}, Deffuant--Weisbuch \cite{De00} as well as the models introduced by Toccaceli et al. \cite{To20}. 

For this purpose, we will use two different metrics to assess the shift in the opinions of normal agents:  the mean/standard deviation of final opinion and the interval defined by \textit{primary clusters}. The former takes into account all the opinions of the population, although it does not adequately distinguish the clustering. 

The latter considers only the range of the largest clusters, allowing the elimination of isolated opinions or small clusters at the extremes of the range of opinions. This second metric \cite{Ga24} consists in assigning an \textit{effective} weight \cite{La79} to each cluster of the final opinion distribution and discarding those which weight is smaller than a \textit{screening value} $\delta>0$. The resulting interval containing non--discarded clusters defines a central point and a width, that we will call \textit{center} and \textit{amplitude} respectively, which we use to compute the influence of manipulators.  

Following \cite{Ga24}, in order to calculate the effective weights, we will smooth the final distribution by Gaussian kernel smoothers 
\[
G_k(x,r_k)=e^{-\left(\frac{x-r_k}{\alpha \varepsilon}\right)^2}
\]
where $\varepsilon$ will be the confidence threshold and $r_k=-1+\frac{2k-1}{h}$ are the middle points of the intervals $[\frac{-h+2(k-1)}{h},\frac{-h+2k}{h}]$ for $k=1,\ldots ,h$. Then, the value of the smooth distribution  $S$ at $r_k$ is 
\[
S(r_k)=\sum_{j=1}^{N}G_k(r_k,x_j)  .
\]
Once we have determined the indexes $\{k_1,\ldots , k_s\}$ such that $S(r_{k_i})$ are local maximums of $S$ for $i=1,\ldots , s$, then the weight of the cluster at $r_{k_i}$ is 
\[
W_{k_i}=\frac{w_{k_i}}{\sum_{i=1}^{s}w^2_{k_i}}
\]
where $w_{k_i}=\frac{S(r_{k_i})}{\sum_{i=1}^{s}S(r_{k_i})}$. Finally, we will discard the clusters $r_{k_i}$ such that $W_{k_i} < \delta$. The remaining clusters, called \textit{primary clusters} and denoted by $\{\overline{r}_1,\ldots ,\overline{r}_b\}$, define an interval $Q=[\mathrm{min}\{\overline{r}_j\}, \mathrm{max}\{\overline{r}_j\}]\subset [-1,1]$ such that its center $\mathbf{c}_b$ and its amplitude $\mathbf{a}_b$ given by 
\[
\left\{
\begin{tabular}{l}
$\mathbf{c}_b = \frac{1}{2}( \mathrm{max}\{\overline{r}_j\} + \mathrm{min}\{\overline{r}_j\})$ \\
$\mathbf{a}_b = \frac{1}{2}( \mathrm{max}\{\overline{r}_j\} - \mathrm{min}\{\overline{r}_j\})$
\end{tabular}
\right.
\]
allow to measure the influence of manipulators. We will implement this method with values $h=200$ and $\alpha = 0.1$.

\bigskip

There are two features of the manipulative agents that we will vary in this article: the number of manipulators $K$ and the speed at which manipulators change their opinion to influence normal agents. One expects that, for a group of manipulators, the greater their number or the slower their speed, the greater the influence they can have on the other agents. 

Notice that throughout this article, we will consider the confidence thresholds of all agents to be equal, that is $\varepsilon_i=\varepsilon$ for all $i=1,\ldots , N$.

\section{The Hegselmann--Krause model}\label{sec:HKclassic}

The first case we will study is the \emph{Hegselmann--Krause model} \cite{He02} or HK--model for short. For the agent $p_i$, we consider the set $\Omega_i$ to be the whole confidence set $J_i$ and the new opinion $x_i$ of the agent $p_i$ is just the arithmetic mean values of all agents in $J_i$. So, we get 
\begin{equation}\label{eq-opinion-HK}
x_i(t+1) = \frac{1}{\vert J_{i}(t)\vert}\sum_{j\in J_i(t)}x_j(t) 
\end{equation}
regarding that $i\in J_i$ for all $i$ and then, if $J_i(t) = \{ i \}$ we have that $x_i(t+1) = x_i(t)$.

\begin{figure}[h]
\centering
 \includegraphics[scale=0.6]{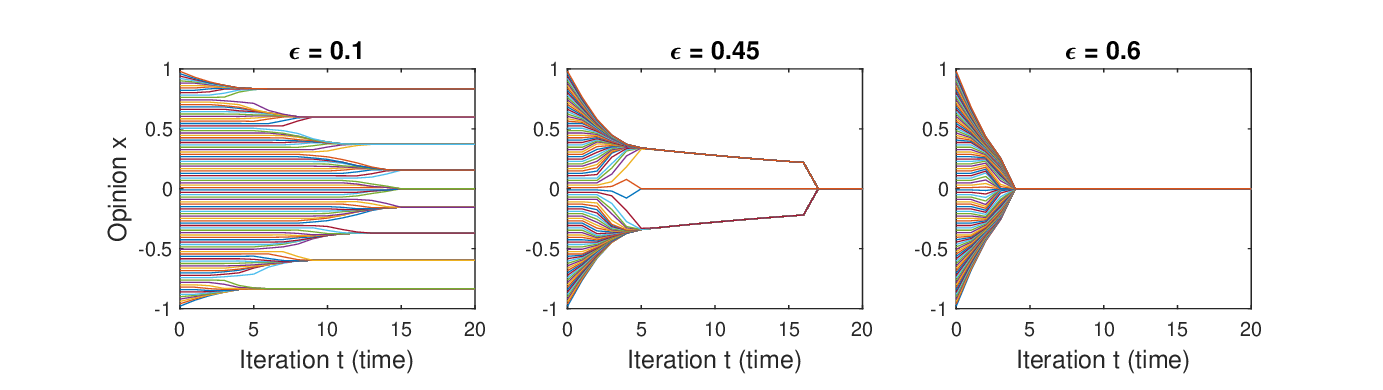}
 \caption{Simultaneous HK--model: opinion dynamics for an initial distribution of 100 equispaced points $x_i(0)=-1+\frac{2i}{101}$ for $i=1,\ldots,100$. The constant values of $\varepsilon$ are specified above every graph.}
  \label{figure-scBC1rev1}
\end{figure}

To illustrate the opinion dynamics in this model, in figure \ref{figure-scBC1rev1}, we consider three simulations for an equispaced initial distribution $\{x_i(0)\}_{i=1}^{N}\subset [-1,1]$ for $N=100$ agents, with different constant confidence thresholds $\varepsilon\in\{0.1,0.45,0.6\}$ and $T=20$ iterations. Observe that, for small values of $\varepsilon$, clusters appear at the final distribution  $\{x_i(T)\}_{i=1}^{N}$ whereas for larger values of $\varepsilon$, the number of cluster decreases to 1, that is, consensus is reached. Note also that for intermediate values of $\varepsilon$, some clusters may appear that are not stable over time and end up grouping, reaching a consensus, after a relatively large number of iterations (case $\varepsilon=0.45$ in the center of figure \ref{figure-scBC1rev1}), see \cite{He02}.

\begin{figure}[h]
\centering
 \includegraphics[scale=0.6]{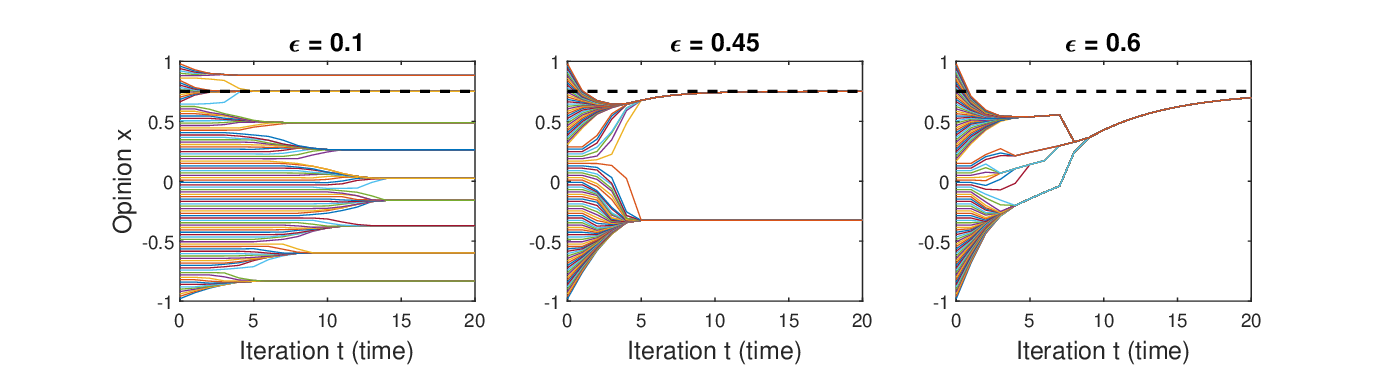}
 \caption{Simultaneous HK--model: opinion dynamics for the same initial distribution as figure \ref{figure-scBC1rev1} with one group of 20 stubborn agents each with opinion at $x=0.75$ (marked with a dashed line).}
  \label{figure-scBCRad1rev1}
\end{figure}

Figure \ref{figure-scBCRad1rev1} shows the opinion dynamics for the equispaced initial distribution $\{x_i(0)\}_{i=1}^{N}\subset [-1,1]$ with $N=100$, plus a group of 20 additional stubborn agents with opinion $x=0.75$. It is clear that the existence of stubborn groups change the dynamical behaviour of the opinions depending, among others, on the size of the group.

\subsection{The HK--model with manipulative agents}\label{sec:HKmodel}

Consider an agent space $P = \{p_i\}_{i=1}^{N}$ with distribution of opinion at time $t$ given by $\{x_i(t)\}_{i=1}^{N}$ and $K$ manipulative agents with opinions $\{f_j(t)\}_{j=1}^{K}$ at time $t$. 
Then, the dynamics of this model is ruled by 
\begin{equation}\label{eq-BCMan-opinion-dynamic}
x_i(t+1) = \frac{1}{\vert \overline{J}_{i}(t)\vert }\sum_{k\in \overline{J}_i(t)}z_k(t) = \frac{1}{\vert J_{i}(t)\vert + \underset{k>N}{\sum_{k\in \overline{J}_i(t)}}1}\left[\underset{k\leq N}{\sum_{k\in \overline{J}_i(t)}}x_k(t) + \underset{k> N}{\sum_{k\in \overline{J}_i(t)}} f_k(t)\right]
\end{equation}
according to (\ref{eq-zk-definition}), where $\overline{J}_i(t)$ is defined as in (\ref{eq-J-manipulators}) and $J_i(t)$ as in (\ref{eq-set-of-confidence}).

\begin{figure}[h]
\centering
 \includegraphics[scale=0.6]{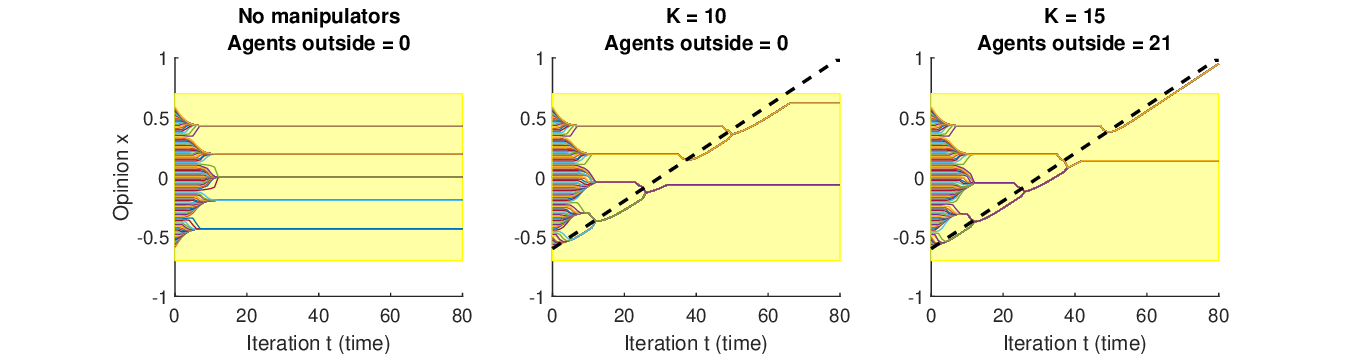}
 \caption{Simultaneous HK--model: dynamics of an equispaced initial opinion distribution of 100 normal agents in $[-0.6,0.6]$ with a confidence threshold $\varepsilon=0.1$. The pictures represent the opinion evolution without manipulative agents (\textit{left}), with a group of $K=10$ manipulators (\textit{center}) and a group of $K=15$ manipulator (\textit{right}). The manipulative group changes its opinion linearly from $f(0)=-0.6$ to $f(80)=1$. The yellow area represents the initial Overton window and the dashed line the trajectory of the opinion of the manipulative group.}
  \label{figure-scBCOver}
\end{figure}

As mentioned in section \ref{sec:Overtonw} above, we can use the HK--model to illustrate the shift of the Overton window.
In figure \ref{figure-scBCOver}, we show three simulations in the opinion space $[-1,1]$ in which we consider the interval $[-0.7,0.7]$ as the Overton window. We choose an equispaced initial distribution of $N=100$ normal agents with acceptable opinions within the interval $[-0.6,0.6]$ and fix the confidence threshold to $\varepsilon=0.1$. The yellow area represents the Overton window along the time $t>0$. 
The graphic on the left describes the evolution of opinions without manipulative agents. In the other graphics, one group of $K=10$ (\textit{center}) and $K=15$ (\textit{right}) manipulators respectively, intervenes in the iteration. The opinion of all manipulative agents moves linearly from $f(0)=-0.6$ to $f(80)=1$. Note that in the second and third cases, the range of opinions shifts upwards to $[-0.06,0.62]$ and $[0.13,1]$ respectively. Therefore, the Overton windows must also have shifted in the same direction. The graph on the right shows that the manipulators are able to push the opinion of 21 normal agents to the extreme $x=1$ of the opinion space, outside the initial Overton window. 

\subsubsection{Analysis of the opinion dynamics between manipulative and normal groups}\label{sec:analysisHK}

The effect of manipulators on normal agents in the HK--model can be explained by the analytical relationship between the parameters involved and this relationship also helps understanding in similar models.

Notice, as shown in figures \ref{figure-scBC1rev1}, \ref{figure-scBCRad1rev1} and \ref{figure-scBCOver}, that clusters take shape in each confidence interval after a few iterations, so it is appropriate to analyse the interaction among groups of normal agents and groups of manipulators. 

Let us consider $m$ manipulative groups with opinion $\{f_j(t)\}$ with $K_j\in \mathbb{N}$ agents each for $j=1,\ldots,m$ interacting with $n$ groups with $N_i\in \mathbb{N}$ normal agents with opinion $\{x_{i}(t)\}$ for $i=1,\ldots,n$ in the same confidence interval then, by iteration (\ref{eq-BCMan-opinion-dynamic}) of simultaneous HK--model, for any $\alpha=1,\ldots ,n$ we have that the opinion of an agent of the normal group $i=\alpha$ is modified by
\begin{equation}\label{eq-analysisHK-multiple}
x_{\alpha}(t+1) = \frac{\sum_{j=1}^{m}K_j f_j(t) + \sum_{i=1}^{n}N_i x_{i}(t)}{\sum_{j=1}^{m}K_j + \sum_{i=1}^{n}N_i}   .
\end{equation}
Since, in simultaneous way, every agent in the confidence interval modifies its opinion under the iteration, the right hand side of (\ref{eq-analysisHK-multiple}) is the same for all $\alpha$, so it does not depend on $\alpha$, then all normal groups reach a local consensus in the first iteration becoming an aggregated normal group of $N=\sum_{i=1}^{n}N_i$ agents with the same opinion. 
Moreover, by the linearity of the mean values, we can simplify the study of the interaction among several groups of normal agents and manipulators to the interaction of just one group of each type. The opinions of such groups will be the mean values of the opinions of all agents in each one. 
Note that these new groups will be virtual because there may not be real agents with its opinion.

So, let us consider one group of $N$ normal agents such that $x_i (0) =x_0$ for $i=1,\ldots, N$ and one group of $K$ manipulative agents such that their opinions are defined by $f(t)=f_0 +\lambda t$ with $\vert f_0 - x_0 \vert < \varepsilon$ and $\lambda\in\mathbb{R}$. Since $x_i (0) =x_j (0)$ for all $i,j=1,\ldots, N$ then we can do without sub-indexes naming the opinion of normal agents by $x=x_i$. Then we have that opinion dynamics follows the formula 
\[
x(t+1) = \frac{Kf(t) +N x(t)}{K+N}
\]
for $t\geq 0$.

\begin{proposition}\label{prop-slope-m}
If $\vert f(t)-x(t) \vert \leq \varepsilon$ and $\vert \lambda \vert \leq \frac{K \varepsilon}{K+N}$ then $\vert f(t+1)-x(t+1) \vert \leq \varepsilon$.
\end{proposition}

\begin{proof}
Assuming the hypotheses, we have
\begin{align*}
\left| f(t+1)-x(t+1)\right| & = \left| \left(f(t)+ \lambda \right) - \left( \frac{Kf(t) +N x(t)}{K+N} \right) \right| = \\
& = \left| \frac{(K+N)f(t) +\lambda(K+N)- Kf(t)-N x(t)}{K+N} \right| = \\
& = \left| \frac{N\left(f(t)-x(t)\right) +\lambda(K+N) }{K+N} \right| = \\
& = \left| \frac{N}{K+N}\left(f(t)-x(t)\right) + \lambda \right| \leq \\
& \leq \left| \frac{N}{K+N}\left(f(t)-x(t)\right) \right| + \left|  \lambda \right| \leq \\
& \leq \frac{N \varepsilon}{K+N} + \left| \lambda \right| \leq \\
& \leq \frac{N \varepsilon}{K+N} + \frac{K \varepsilon}{K+N} =  \varepsilon
\end{align*} 
as we claimed.
\end{proof}

\begin{proposition}\label{prop-distance}
If $\vert f(t)-x(t) \vert \leq \varepsilon$ for all $t\in\{0\}\cup \mathbb{N}$, then we have
\begin{equation}\label{eq-distance}
f(t)-x(t) = \left( \frac{N}{K+N} \right)^t \left(f_0 -x_0\right) + \frac{\lambda(K+N)}{K}\left(1-\left( \frac{N}{K+N} \right)^t\right)   
\end{equation}
for all $t\in\{0\}\cup \mathbb{N}$.
\end{proposition}

\begin{proof}
By hypothesis, the opinion of the group of normal agents is under the influence of the group of manipulators for all $t\in\{0\}\cup \mathbb{N}$. So, we can prove the result by induction. Indeed, the formula (\ref{eq-distance}) is trivially verified for $t=0$. Assuming equality (\ref{eq-distance}) is true for $t$, we can write that 
\begin{align*}
f(t+1)-x(t+1) & = \left(f(t)+ \lambda \right) - \left( \frac{Kf(t) +N x(t)}{K+N} \right) = \\
& = \frac{(K+N)f(t) +\lambda(K+N)- Kf(t)-N x(t)}{K+N} = \\
& = \frac{N\left(f(t)-x(t)\right) +\lambda(K+N) }{K+N} = \\
& = \frac{N\left[\left(\frac{N}{K+N} \right)^t \left(f_0 -x_0\right) + \frac{\lambda(K+N)}{K}\left(1-\left( \frac{N}{K+N} \right)^t\right) \right] +\lambda(K+N)}{K+N}  = \\
& = \left(\frac{N}{K+N} \right)^{t+1} \left(f_0 -x_0\right) + \frac{\lambda N}{K} + \lambda -\frac{\lambda(K+N)}{K}\left( \frac{N}{K+N} \right)^{t+1}  = \\
& = \left( \frac{N}{K+N} \right)^{t+1} \left(f_0 -x_0\right) + \frac{\lambda(K+N)}{K}\left(1-\left( \frac{N}{K+N} \right)^{t+1}\right)
\end{align*}
so, the formula (\ref{eq-distance}) is also true for $t+1$ and, by the hypothesis of induction, it is satisfied for all positive integer $t\in\{0\}\cup \mathbb{N}$.
\end{proof}

A trivial consequence of propositions \ref{prop-slope-m} and \ref{prop-distance} is the following corollary.

\begin{corollary}\label{cor-slope-m}
The following conditions 
\begin{enumerate}
\item\label{cor-cond-1} $\vert f_0-x_0 \vert \leq \varepsilon$ and $\vert \lambda \vert \leq \frac{K \varepsilon}{K+N}$. \vspace{5pt}
\item\label{cor-cond-2} $\vert f(t)-x(t) \vert \leq \varepsilon$ for all $t\in\{0\}\cup \mathbb{N}$.
\end{enumerate}
are equivalent.
\end{corollary}

\begin{proof}
(\ref{cor-cond-1}) $\Rightarrow$ (\ref{cor-cond-2}). It is trivial by mathematical induction using proposition \ref{prop-slope-m}.

(\ref{cor-cond-2}) $\Rightarrow$ (\ref{cor-cond-1}). By equation (\ref{eq-distance}), we have 
\[
\left| f(t)-x(t) \right| = \left| \left( \frac{N}{K+N} \right)^t \left(f_0 -x_0\right) + \frac{\lambda(K+N)}{K}\left(1-\left( \frac{N}{K+N} \right)^t\right) \right|  
\]
and computing limits 
\[
\lim_{t\mapsto \infty} \left| \left( \frac{N}{K+N} \right)^t \left(f_0 -x_0\right) + \frac{\lambda(K+N)}{K}\left(1-\left( \frac{N}{K+N} \right)^t\right) \right| \leq \varepsilon
\]
we obtain
\[
\left| \frac{\lambda(K+N)}{K} \right| \leq \varepsilon \qquad \Longrightarrow \qquad  \left| \lambda \right| \leq \frac{K \varepsilon}{K+N}
\]
as claimed.
\end{proof}

The condition (\ref{cor-cond-1}) of corollary \ref{cor-slope-m}, that is,
\begin{equation}\label{eq-condition-one-group}
  \left| \lambda \right| \leq \frac{K \varepsilon}{K+N} \quad  \text{ with } \left| f_0-x_0 \right| \leq \varepsilon
\end{equation}
gives us the relationship among the quantities $K$, $N$, $\lambda$ and $\varepsilon$ to ensure that the group of manipulators influences the group of normal agents for all $t>0$. Observe that $\lambda$ describes the speed of change of the opinion of the manipulative group, so when this rate decreases, the ability of the manipulators to influence normal agents increases. Similarly, when the proportion of manipulative agents to the total number of agents (manipulators and normals) is increased, which is given by $w_f = \frac{K}{K+N}$, the ability of the manipulative group to influence the normal population is also increased. The same effect has the confidence threshold $\varepsilon$ so, in the HK--model, the more distrustful the population (small values of $\varepsilon$) the lower the influence of the manipulative group. If each normal agent trusts any other opinion (large values of $\varepsilon$), the manipulative group will attract the opinion of the group (or groups) of normal agents to its own, thus achieving the manipulation of the whole society.

\begin{figure}[h]
\centering
 \includegraphics[scale=0.6]{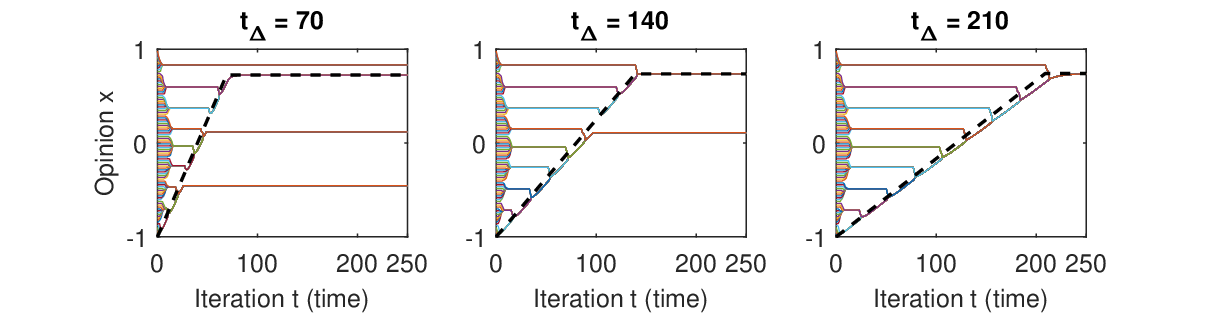}
 \caption{Simultaneous HK--model: initial opinion distribution of 100 normal agents in $[-1,1]$ with a confidence threshold $\varepsilon=0.1$ and a group of 8 manipulators changing their opinion linearly from $f(0)=-1$ to $f(t_{\Delta})=0.75$ for $t_{\Delta}\in\{70,140,210\}$ iterations keeping it constant for $t\geq t_{\Delta}$. The dashed line is the trajectory of the opinion of the manipulative group.}
  \label{figure-scBCMan2rev1A}
\end{figure}

\begin{figure}[h]
\centering
 \includegraphics[scale=0.6]{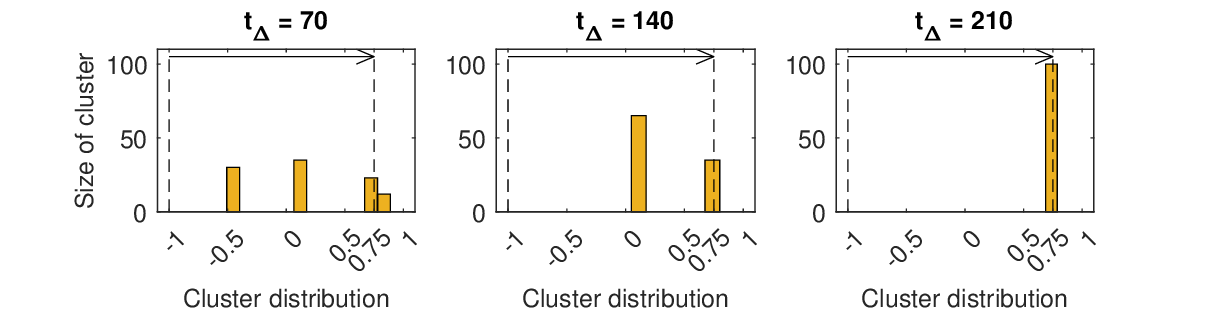}
 \caption{Simultaneous HK--model: size of the cluster in the final distribution of the same simulation of figure \ref{figure-scBCMan2rev1A}. The manipulators groups the opinions of neighbouring normal agents along its way until the cluster is big enough not to satisfy the condition (\ref{eq-condition-one-group}).}
  \label{figure-scBCMan2rev1B}
\end{figure}

Corollary \ref{cor-slope-m} implies that, whenever $\left| \lambda \right| > \frac{K \varepsilon}{K+N}$ with $\left| f(t_0)-x(t_0) \right| \leq \varepsilon$, there will exist $T\geq 1$ such that $\left| f(t_0+T)-x(t_0+T) \right| > \varepsilon$.
Then, the influence of manipulators stops at $t=t_0+T$ leaving the confidence interval of the normal group. This feature is shown in figures \ref{figure-scBCMan2rev1A} and \ref{figure-scBCMan2rev1B}, in which the manipulative group moves its opinion from $f(0)=-1$ to $f(t_{\Delta})=0.75$ in $t_{\Delta}\in\{70,140,210\}$ iterations before it becomes stubborn. So, $\lambda\in\{0.025 , 0.0125 , 0.0083\}$ and the manipulators attract normal agents into a cluster which follows the trajectory of opinion of manipulators until the number $N$ of normal agents is big enough not to verify the condition $\left| \lambda \right| \leq \frac{K \varepsilon}{K+N}$. 
Thus, the manipulative group leaves the confidence interval of the cluster of normal agents whose opinion is not longer modified because there is not another agent in said interval.

\subsubsection{Simulations for HK--model with one manipulative group}\label{sec:HKmodelsim}

In order to study the behaviour of the model with an only group of manipulators, we will compare the central points and widths of the final distributions using the metrics of the mean/standard deviation and the interval of primary clusters as a function of the number $K$ of agents in the manipulative group and the speed $\lambda$ at which they change their opinion. 

Note that, since the ratio of the variation $\lambda$ of the manipulators' opinion  and the number of iterations $t_{\Delta}$ in which the manipulative group change its opinion are related by $\lambda=\frac{f(t_{\Delta})-f(0)}{t_{\Delta}}$, we will use the variable $t_{\Delta}$ instead of $\lambda$.

Fixed the confidence threshold to $\varepsilon = 0.1$, we consider an equispaced initial distribution $\mathbf{x}(0)=\{-1+\frac{2i}{101}\}_{i=1}^{100}$ of normal agents and a group of $K$ manipulators whose opinion changes linearly from $f(0)=-1$ to $f(t_{\Delta})=1$ and then $f(t)=1$ for $t\geq t_{\Delta}$. The values of $K$ and $t_{\Delta}$ are chosen to be $K=0,\ldots, 30$ and $t_{\Delta}\in \{10j\}_{j=0,\ldots 30}$. The criterion for stopping iterations is 
\[
\underset{j}{\mathrm{max}}\vert x_j(t-1)-x_j(t) \vert \leq 5\cdot 10^{-4}   .
\]

\begin{figure}[h]
\centering
 \includegraphics[scale=0.65]{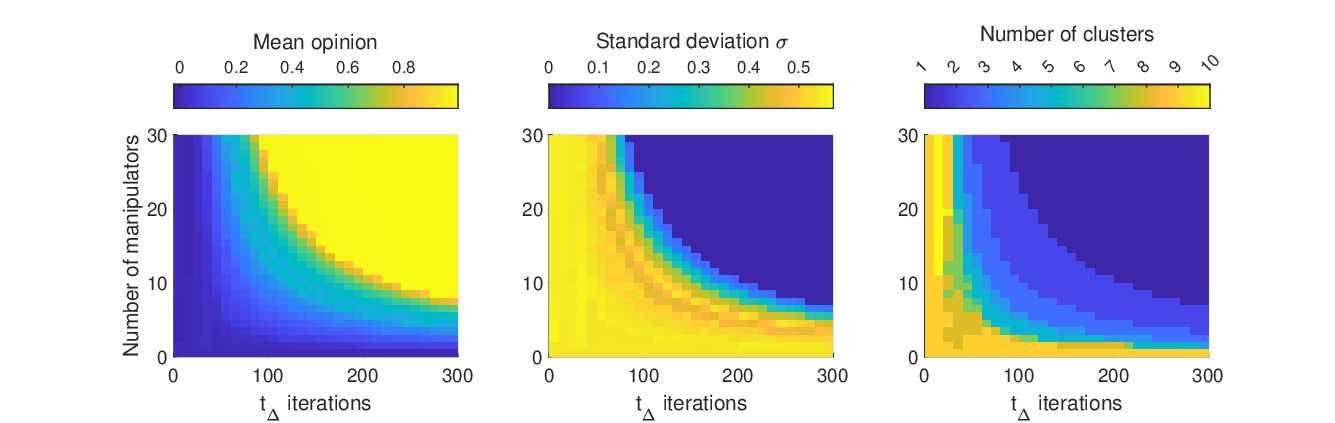}
 \caption{Simultaneous HK--model: mean value (\textit{left}), standard deviation (\textit{center}) and number of final clusters (\textit{right}) depending on the number $K$ of manipulators and the number $t_{\Delta}$ of iterations the manipulators are using to reach their final opinion $f(t_{\Delta})=1$ from their initial opinion $f(0)=-1$. The confidence threshold is fixed at $\varepsilon=0.1$ constant.}
  \label{figure-scBCMan1pend1rev1A}
\end{figure}

In figure \ref{figure-scBCMan1pend1rev1A}, we represent the mean value, standard deviation and number of clusters (maybe isolated opinions) of the final distribution of normal opinions $\mathbf{x}$ by a color map depending on the number of iterations $t_{\Delta}$ and the number $K$ of agents in the manipulative group. We observe that, for small values of the variables, the effect of the manipulative group is not very significant and the three values remain close to the achieved without manipulators.

Small values of $t_{\Delta}$ imply high speeds in the change of opinion of the manipulators. When the values of $t_{\Delta}$ are increased, the opinion of this group changes more slowly and this allows the manipulators to shift the opinions of normal agents in the same direction as the opinion of the manipulative group. In these cases the mean values are close to the final opinion of the manipulators (bright yellow in the left graph) meanwhile the standard deviations decrease to $\sigma=0$ (dark blue in the center graph), indicating that consensus is reached.

\begin{figure}[h]
\centering
 \includegraphics[scale=0.65]{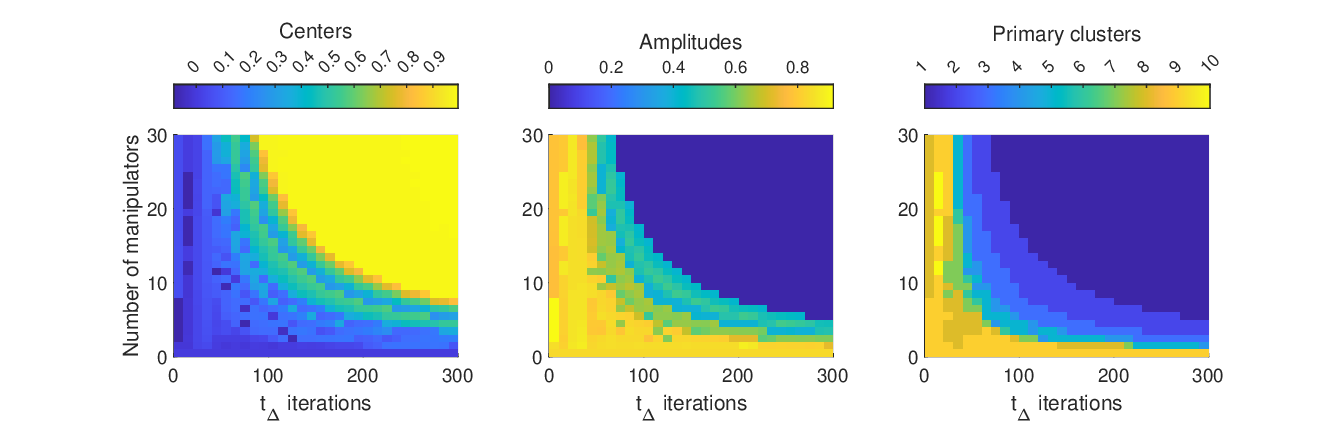}
 \caption{Simultaneous HK--model: features of the interval of primary clusters. Center (\textit{left}), amplitude (\textit{center}) and number of primary clusters obtained in the same simulation as in figure \ref{figure-scBCMan1pend1rev1A}.}
  \label{figure-scBCMan1pend1rev1B}
\end{figure}

\begin{figure}[h]
\centering
 \includegraphics[scale=0.55]{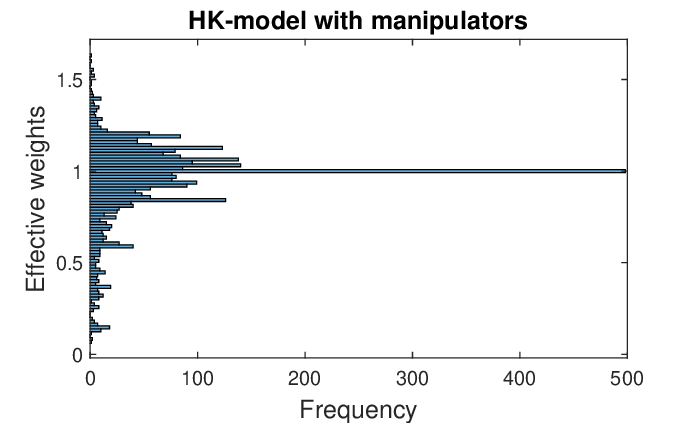}
 \caption{Simultaneous HK--model: histogram of effective weights in simulation of figures \ref{figure-scBCMan1pend1rev1A} and \ref{figure-scBCMan1pend1rev1B}.}
  \label{figure-scBCMan1pend1rev1C}
\end{figure}

In figure \ref{figure-scBCMan1pend1rev1B}, we can observe the results of the same simulation when we consider the center, amplitude and number of primary clusters for a screening value $\delta=0.5$ (see figure \ref{figure-scBCMan1pend1rev1C} for frequency of effective weights). A similar behaviour is obtained but the areas of intermediate values of the center and amplitude are slightly wider than for the mean and standard deviation values at the figure \ref{figure-scBCMan1pend1rev1A}.

Thus, it is noted that, when there is intervention of a manipulative group, the central values of the opinion (mean and center) of the normal population move in the same direction as the manipulators (from $\overline{\mathbf{x}}=0$ without manipulators) while the standard deviations and amplitude decrease. This tells us that the manipulators manage to shift and narrow Overton's window on the subject of study. The narrowing in the range of opinions is also due to the nature of the HK--model, since it tends to concentrate opinions on intermediate values.

%%%%%%%%%%%%%%%%%%%%%%%

\section{The Deffuant--Weisbuch model}\label{sec:DefWeiMod}

The following model we will study is the \emph{Deffuant--Weisbuch model} \cite{De00}, or DW--model for short.

In this model, for each agent $p_i$ we choose the set $\Omega_i=\{i,j\}$ where $j\in \{1,\ldots,n\}$ is a random index such that $j\neq i$, then we consider the iteration  
\begin{equation}\label{eq-opinion-DW}
x_i(t+1) = \left\{
\begin{array}{lcl}
 \displaystyle{\frac{x_i(t)+x_j(t)}{2} }& & \text{ if } \left|x_i(t)-x_j(t)\right|\leq \varepsilon \\
 x_i(t) & & \text{ otherwise }
\end{array}
\right.
\end{equation}

Again, we will implement this model in simultaneous way, i.e., each agent performs the iteration (\ref{eq-opinion-DW}) for each $t$.

\begin{figure}[h]
\centering
 \includegraphics[scale=0.6]{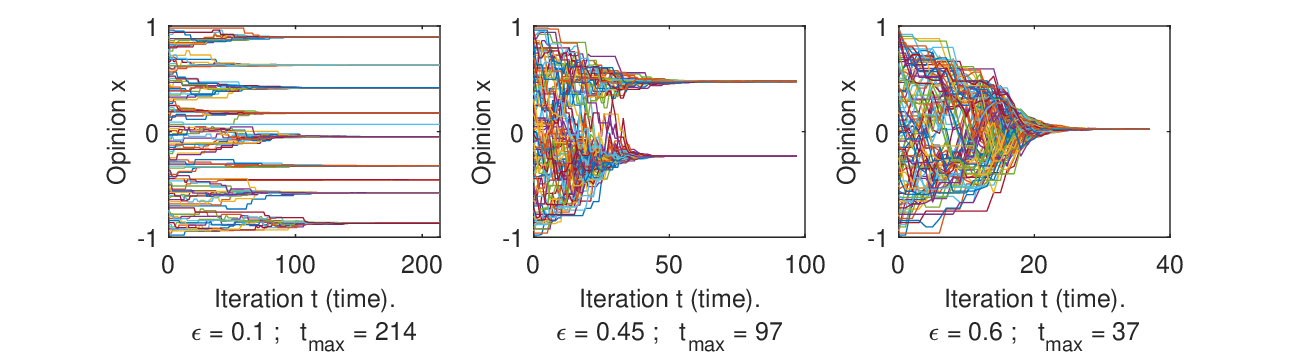}
 \caption{Simultaneous DW--model: opinion dynamics for an initial distribution of $N=100$ equispaced points $x_i(0)=-1+\frac{2i}{101}\in [-1,1]$ with $\varepsilon_i=\varepsilon\in\{0.1,0.45,0.6\}$ constant for all $i=1,\ldots,100$.}
  \label{figure-scDWnrev1}
\end{figure}

In figure \ref{figure-scDWnrev1}, examples of opinion dynamics for a simultaneous DW--model are shown. The distribution of opinions is considered stable whenever the values in each cluster differ less than $5\cdot 10^{-4}$ units achieved in $t_{max}$ iterations.

\begin{figure}[h]
\centering
 \includegraphics[scale=0.6]{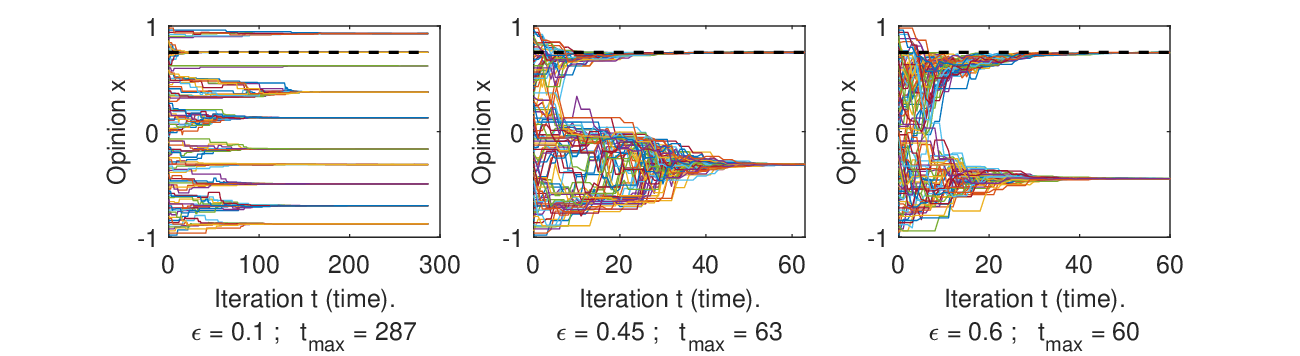}
 \caption{Simultaneous DW--model: opinion dynamics for the same parameters as in figure \ref{figure-scDWnrev1} with a group of 20 stubborn agents with opinion $x=0.75$. The opinion of the stubborn group is shown as a dashed line.}
  \label{figure-scDWRad1rev1}
\end{figure}

If a distribution of opinion interacts with a group of stubborn agents with constant opinion $x(t)=c\in [-1,1]$ for all $t\geq 0$, then the opinion of agents in the confidence interval of the stubborn agents converges to this group's opinion.  A simulation of this situation can be seen in figure \ref{figure-scDWRad1rev1} for the same initial distribution as in figure \ref{figure-scDWnrev1} and a group of 20 stubborn agents (marked by a dashed line) with constant opinion $x(t)=0.75$ for all $t\geq 0$. Observe that the stubborn group attracts the neighbouring normal agents to its opinion.

\subsection{The DW-model with manipulative agents}\label{sec:DWmodel}

Using the notation introduced in (\ref{eq-zk-definition}), we can define the iteration for the DW--model with manipulators by 
\[
x_i(t+1)=\left\{ 
\begin{array}{lcl}
\displaystyle{\frac{x_i(t)+z_j(t)}{2}} & & \vert x_i(t)-z_j(t) \vert \leq \varepsilon \\
x_i(t) & & \vert x_i(t)-z_j(t) \vert > \varepsilon
\end{array}
\right.
\]
where $z_j$ is the opinion of a random agent (normal or manipulator).

\begin{figure}[h]
\centering
 \includegraphics[scale=0.6]{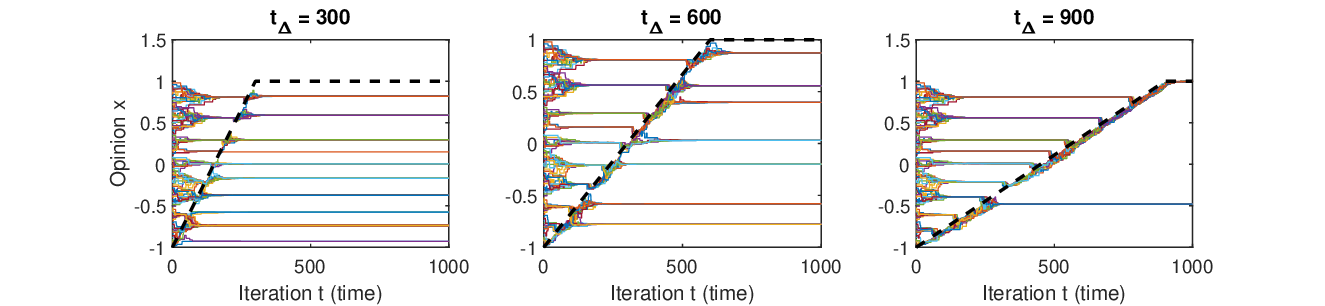}
 \caption{Simultaneous DW--model: trajectories of an equispaced initial opinion distribution of 100 normal agents and a group of 8 manipulators with a confidence threshold $\varepsilon=0.1$. The manipulative group changes its opinion linearly from $f(0)=-1$ to $f(t_{\Delta})=1$ for $t_{\Delta}\in\{300k\}_{k=1}^{3}$ iterations. The dashed line represents the trajectory of the opinion of the manipulative group.}
  \label{figure-scDWMan1rev1A}
\end{figure}

The simulation of figure \ref{figure-scDWMan1rev1A} corresponds to the DW--model with 100 equispaced normal agents, one group of $K=8$ manipulators changing theirs opinions from $f(0)=-1$ to $f(t_{\Delta})=1$ for $t_{\Delta}\in\{300,600,900\}$ and $\epsilon=0.1$.
Note that we obtain similar dynamics to the HK--model but with more fragmentation in the final clusters.

\subsubsection{Simulations for DW--model with one manipulative group}\label{sec:DWmodelsim}

Next, we will study the DW--model in a similar way as we have done in subsection \ref{sec:HKmodelsim}, but in this case, due to its random component, we will perform 100 simulations for each pair of fixed values $(t_{\Delta},K)$, calculating the corresponding central points, dispersion and number of clusters. 
\begin{figure}[h]
\centering
 \includegraphics[scale=0.65]{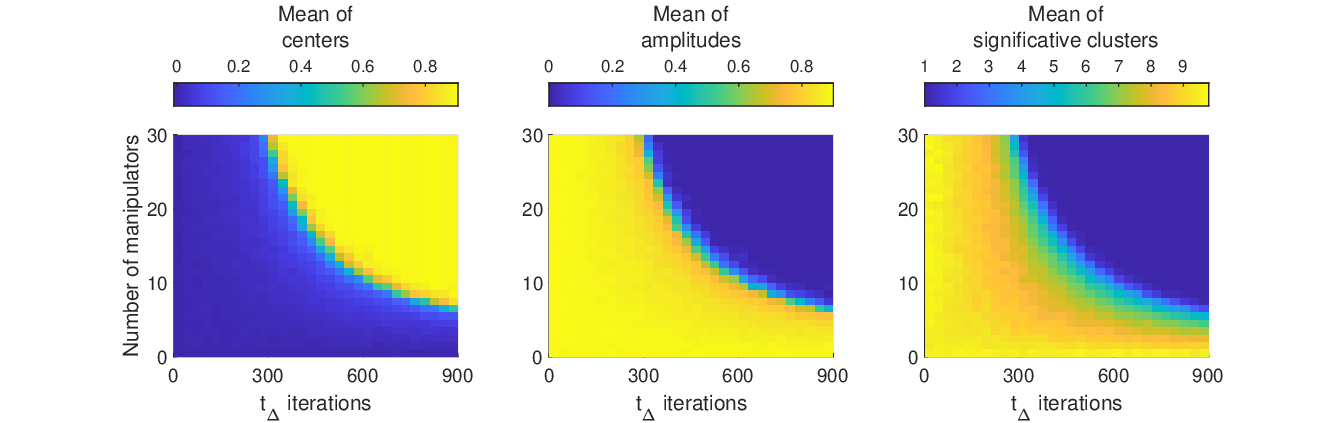}
 \caption{Simultaneous DW--model: averages of mean values (\textit{left}), standard deviations (\textit{center}) and numbers of final clusters (\textit{right}) after 100 simulations for every number $K$ of manipulators in the group and every number $t_{\Delta}$ of iterations, with a constant confidence threshold $\varepsilon=0.1$. The opinion of manipulative group varies linearly from $f(0)=-1$ to $f(t_{\Delta})=1$ before to be stubborn $f(t)=1$ for $t\geq t_{\Delta}$.}
  \label{figure-scDWMan1pend1rev2A}
\end{figure}

The initial distribution of normal agents for all simulations is $\mathbf{x}(0)=\{-1+\frac{2i}{101}\}_{i=1}^{100}$ and, again, we consider a group of $K$ manipulators whose opinion changes linearly from $f(0)=-1$ to $f(t_{\Delta})=1$ remaining constant $f(t)=1$ for $t\geq t_{\Delta}$. The range of $K$ will be the same as for the HK-model: $K\in\{0,\ldots, 30\}$ but the values of $t_{\Delta}$ are chosen higher since the convergence is slower. So, we will take $t_{\Delta}\in \{30j\}_{j=0}^{30}$ and set the stopping criterion of iterations by making it so that, after each iteration, the values of the opinions, rounded to 3 decimal places, are clustered at a distance among clusters greater than the confidence threshold $\varepsilon=0.1$. 

\begin{figure}[h]
\centering
 \includegraphics[scale=0.65]{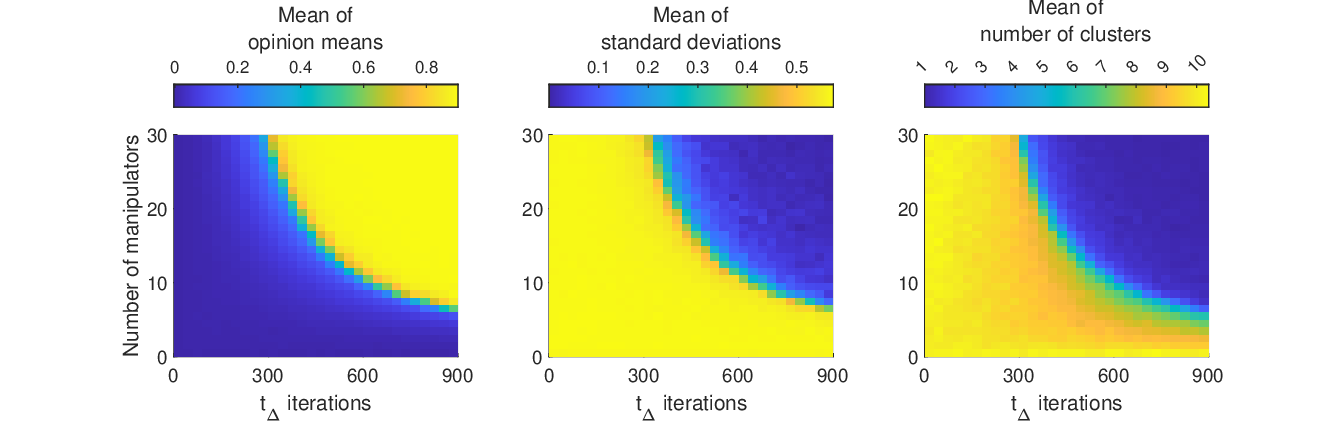}
 \caption{Simultaneous DW--model: averages of centers (\textit{left}), amplitudes (\textit{center}) and number of primary clusters (\textit{right}) after 100 simulations with the same parameters as in figure \ref{figure-scDWMan1pend1rev2A}.}
  \label{figure-scDWMan1pend1rev2B}
\end{figure}

\begin{figure}[h]
\centering
 \includegraphics[scale=0.55]{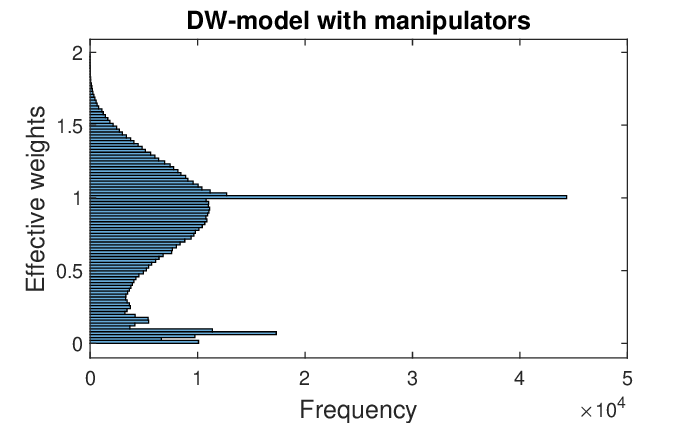}
 \caption{Simultaneous DW--model: histogram of effective weights in simulation of figures \ref{figure-scDWMan1pend1rev2A} and \ref{figure-scDWMan1pend1rev2B}.}
  \label{figure-scDWMan1pend1rev2C}
\end{figure}

In figures \ref{figure-scDWMan1pend1rev2A} and  \ref{figure-scDWMan1pend1rev2B}, it is observed a very similar result that in the case of HK--model except that the transition region of intermediate values (in greens) in each graph is considerably narrower. This implies that when the manipulative group has significant influence, it manages to attract all normal opinions and achieves the consensus in its own opinion. When $t_{\Delta}\lesssim 300$ or $K\lesssim 8$, the normal population shows some resistance to the influence of manipulators, but when manipulators manage to break that resistance, they get consensus in their opinion.
In this case, after observing the effective weights obtained in the simulation (see figure \ref{figure-scDWMan1pend1rev2C}), we have used the screening value $\delta=0.2$ to select the primary clusters.

\section{The Weighted HK--models}\label{sec:WHKmodel}

In order to build an agent--based model to study the influence of fake news on society, Toccaceli et al. \cite{To20}, Milli \cite{Mi21} define different models as in (\ref{eq-opinion-dynamic-general}) comprising mechanisms of attraction and/or repulsion that we will introduce below. This models include a weight function of the neighbouring opinions \cite{Fl17}.

Let us consider an agent $p_i$ and a random opinion $x_j(t)$ in the distribution chosen such that $j\neq i$. We will call \emph{Attractive Weighted HK model} to the iteration (\ref{eq-opinion-AWHKG}) corresponding to the attractive mechanism. The opinion $x_i(t)$ is modified by 
\begin{equation}\label{eq-opinion-AWHKG}
x_i(t+1) = \left\{
\begin{array}{lcl}
\displaystyle{x_i(t) - s_{ij}(t)\frac{\vert x_i(t) + w_{ij}x_j(t)\vert}{2} (1-\vert x_i(t) \vert) } & & \text{ if } \left|x_i(t)-x_j(t)\right|\leq \varepsilon \\
\displaystyle{x_i(t)} & & \text{ if } \left|x_i(t)-x_j(t)\right|> \varepsilon
\end{array}
\right.
\end{equation}
where $s_{ij}(t)=sign(x_i(t)-x_j(t))$. 
The components of the matrix of weights $w=(w_{ij})$ describe the trustworthiness the agent $p_i$ attributes to the $p_j$'s opinion for $j\in \{ j:\vert x_{i}(t)-x_{j}(t) \vert \leq \varepsilon \text{ and } j\neq i\}$ and their values will be taken as $w_{ij}\in[0,1]$.

For the model (\ref{eq-opinion-AWHKG}), if the opinion $x_j(t)$ is contained in the confidence interval of $x_i(t)$, then there is an attractive influence of $x_j(t)$ on the opinion $x_i(t)$, that is, the new opinion $x_i(t+1)$ moves closer to $x_j(t)$. 

The repulsive mechanism, defining the \emph{Repulsive Weighted HK model}, is given by
\begin{equation}\label{eq-opinion-RWHKG}
x_i(t+1) = \left\{
\begin{array}{lcl}
\displaystyle{x_i(t) } & & \text{ if } \left|x_i(t)-x_j(t)\right|\leq \varepsilon \\
\displaystyle{x_i(t) + s_{ij}(t)\frac{\vert x_i(t) + w_{ij}x_j(t)\vert}{2} (1-\vert x_i(t) \vert) } & & \text{ if } \left|x_i(t)-x_j(t)\right|> \varepsilon
\end{array}
\right.
\end{equation}
where, again $s_{ij}(t)=sign(x_i(t)-x_j(t))$. Now, if $x_j(t)$ is outside the confidence interval of $x_i(t)$ then $x_i(t+1)$ moves away from $x_j(t)$, so we consider that the influence of $x_j(t)$ on the opinion $x_i(t)$ is repulsive. 

We can combine both mechanisms in one model called \emph{Attractive--Repulsive Weighted HK model}. It is defined by 
\begin{equation}\label{eq-opinion-ARWHKG}
x_i(t+1) = \left\{
\begin{array}{lcl}
\displaystyle{x_i(t) - s_{ij}(t)\frac{\vert x_i(t) + w_{ij}x_j(t)\vert}{2} (1-\vert x_i(t) \vert)} & & \text{ if } \left|x_i(t)-x_j(t)\right|\leq \varepsilon \\
\displaystyle{x_i(t) + s_{ij}(t)\frac{\vert x_i(t) + w_{ij}x_j(t)\vert}{2} (1-\vert x_i(t) \vert) } & & \text{ if } \left|x_i(t)-x_j(t)\right|> \varepsilon
\end{array}
\right.
\end{equation}

Note that the term $\phi_i(t) = 1-\vert x_i(t) \vert$ can be thought of as the probability density function $g(x)=1-\vert x \vert$ that gives weight to the interval $[-1,1]$. This makes moderate opinions (close to $x=0$) more likely to be modified and, therefore, after stabilization of opinion dynamics, most opinions are found in clusters $x=\pm 1$ which are, in fact, fixed points of the iterations. Some opinions may fall outside these radical groups depending on the values of the confidence threshold $\varepsilon$. If the value of $\varepsilon$ is large enough, the final distribution is bimodal, polarized at $x=\pm 1$.

\begin{figure}[h]
\centering
 \includegraphics[scale=0.6]{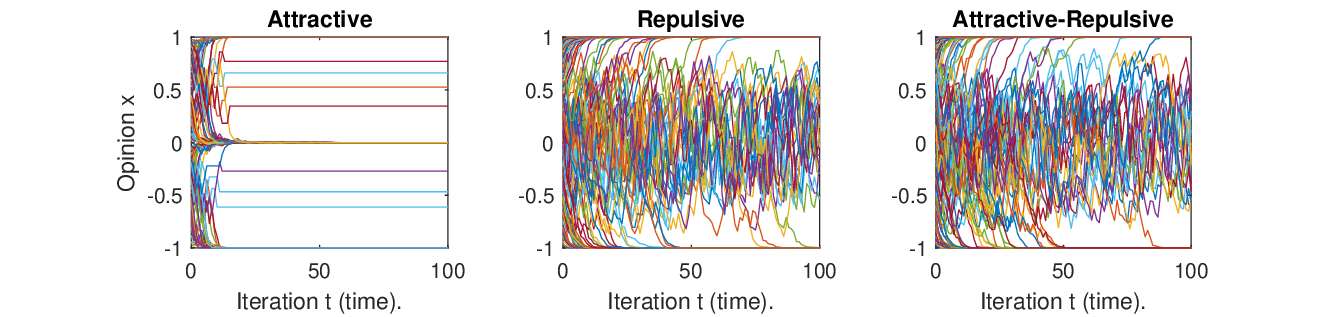}
 \caption{Simultaneous Attractive, Repulsive and Combined Weighted HK models: opinion dynamics for an equispaced initial opinion distribution of $N=100$ normal agents, a threshold of confidence of $\varepsilon=0.1$ and the same random matrix $w=(w_{ij})$.}
  \label{figure-scAtrRepEx2rev1}
\end{figure}

In the figure \ref{figure-scAtrRepEx2rev1} and in order to compare the different dynamics, simulations of 100 iterations for the Simultaneous Attractive, Repulsive and combined (Attractive-Repulsive) models are shown for the same equispaced initial opinion distribution of $N=100$ normal agents with a confidence threshold of $\varepsilon=0.1$ and a random weight matrix $w=(w_{ij})$. When the value of $\varepsilon$ is small, the final state for the Attractive model consists of three big clusters and some individual opinions in between. The largest cluster has the opinion of about 45 normal agents and it is located at $x=0$. There are two smaller clusters located at $x=\pm 1$ with about 24 normal agents each one. The opinions of the rest of normal agents are isolated in between those three clusters. For Repulsive and Attractive--Repulsive models, the central opinions oscillate with no convergence. 

As the value of $\varepsilon$ grows in the Attractive model, the distance between contiguous final opinions becomes larger until isolated opinions converge to one of the major clusters. In addition, for large values of $\varepsilon$, the final state can consist of a single cluster at $x=0$ or the two extreme clusters at $x=\pm 1$.

In the cases of Repulsive and the combined Attractive--Repulsive models, the growth of $\varepsilon$ shows dynamics with a temporal oscillation of some moderate opinions that converge, finally, to $x=\pm 1$ for larger $t$. For values $\varepsilon=0.6$ and larger, these models converge to $x=\pm 1$ without temporal oscillations.

\subsection{The Weighted HK models with manipulative agents}

In this section, we will consider the simultaneous Weighted HK models interacting with one group of manipulators. The iterations in these models are analogous to expressions (\ref{eq-opinion-AWHKG}), (\ref{eq-opinion-RWHKG}) and (\ref{eq-opinion-ARWHKG}) except we have to substitute $x_j$ by $z_j$ where the latter will be an agent (manipulator or normal) in the confidence interval of the normal agent $x_i$.

\subsubsection{Attractive Weighted HK model with manipulators}

The opinion dynamics under Weighted HK models with a manipulative group is richer than for the previous models depending on the values of the threshold $\varepsilon$, the number of iterations $t_{\Delta}$ and the ratio between the number of manipulators in the group and the number of normal agents.

\begin{figure}[h]
\centering
 \includegraphics[scale=0.6]{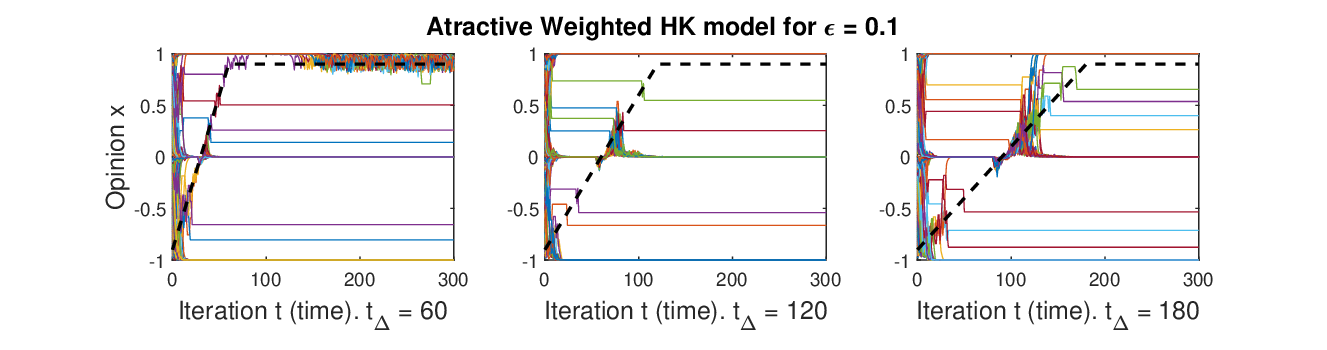}
 \caption{Simultaneous Attractive Weighted HK model: trajectories of 100 normal opinions initially equispaced normal agents and a group of 60 manipulators. The confidence threshold is fixed to $\varepsilon=0.1$ and the iterations $t_{\Delta}\in\{60,120,180\}$. The initial opinion of manipulators is $f(0)=-0.9$ and the final opinion is $f(t_{\Delta})=0.9$. The trajectory of the opinion of manipulative group follows the dashed line.}
  \label{figure-scAtrNMan2Arev1}
\end{figure}

As described in section \ref{sec:WHKmodel} above, whenever the value of $\varepsilon$ is small (see the graphic at the left of figure \ref{figure-scAtrRepEx2rev1}), the final distribution comprises three major clusters at $x=-1,0,+1$ with some few isolated opinions in between. So, the influence of the manipulative group provokes perturbations on the major clusters and can modify the isolated opinions even grouping them to some cluster but it can not change the opinion of a cluster. This behaviour is illustrated in figure \ref{figure-scAtrNMan2Arev1}, where a simulation is done with 100 normal opinions initially equispaced and a manipulative group of 60 agents for a confidence threshold $\varepsilon=0.1$. The number of iterations $t_{\Delta}$ to move the opinion of the manipulators from $f(0)=-0.9$ to $f(t_{\Delta})=0.9$ is chosen as $t_{\Delta}\in\{60,120,180\}$. 

\begin{figure}[h]
\centering
 \includegraphics[scale=0.6]{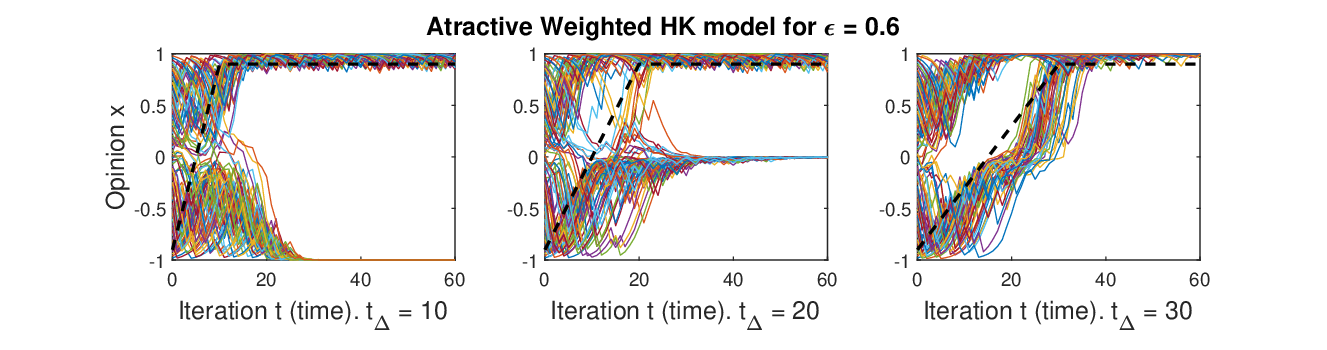}
 \caption{Simultaneous Attractive Weighted HK model: trajectories of 100 normal opinions initially equispaced normal agents and a group of 60 manipulators. The confidence threshold is fixed to $\varepsilon=0.6$ and the iterations $t_{\Delta}\in\{10,20,30\}$. The initial opinion of manipulators is $f(0)=-0.9$ and the final opinion is $f(t_{\Delta})=0.9$. The trajectory of the opinion of manipulative group follows the dashed line.}
  \label{figure-scAtrNMan2Brev1}
\end{figure}

For larger values of $\varepsilon$, a manipulative group can influence the opinion of clustered normal agents.
The simulation of figure \ref{figure-scAtrNMan2Brev1} is made on the same parameters as in figure \ref{figure-scAtrNMan2Arev1} except for a confidence threshold $\varepsilon=0.6$ and $t_{\Delta}\in\{10,20,30\}$. Notice that the final clustered opinions change in each graph: $x=\pm 1$ (left), $x=0,1$ (center) and $x=1$ (right). There are oscillating opinions between $x=0.9$ and $x=1$ because the manipulators become stubborn within the confidence interval of normal agents close to $x=1$.  For $t_{\Delta}=30$ (right), the manipulators move all the normal opinions to an only cluster $x=1$ and since the number of manipulators is lesser than the number of neighbouring normal agents, the manipulative group is not strong enough to cause the opinion of normal agents to oscillate for many iterations between $x=0.9$ and $x=1$.

\subsubsection{Simulations for Attractive Weighted HK model with one manipulative group}

As in section \ref{sec:HKmodelsim}, we will analyse how a manipulative group with $K$ agents affects the opinion dynamics of the simultaneous Attractive Weighted HK model. We will proceed by fixing the confidence threshold to $\varepsilon=0.6$ because the influence of the manipulative group for small values of $\epsilon$ is not significant.

We will consider an initial equispaced distribution $\mathbf{x}(0)=\{-1+\frac{2i}{101}\}_{i=1}^{100}$ of normal agents and one manipulative group with $K\in \{ 5j \}_{j=0}^{40}$ agents which change their opinion from $f(0)=-0.9$ to $f(t_{\Delta})=0.9$ for $t_{\Delta} \in \{ 5j \}_{j=0}^{40}$ and then, for the values $t\in [t_{\Delta},500]$ the manipulative group becomes stubborn.  
For every pair $(t_{\Delta},K)$, we will do 100 simulations and compute the average values obtained for the central point and width in the two used metrics. 
We have to take into account that, as seen in figures \ref{figure-scAtrNMan2Arev1} and \ref{figure-scAtrNMan2Brev1}, we can find cases in which the opinion dynamics does not converge and there exists normal agents with oscillating opinions. So, for each simulation, we will consider the opinion distribution in two particular iterations: $t=t_{\Delta}$ where the manipulative group reaches its final opinion and $t=500$ which allows the stabilization of the distribution in case it occurs. The comparative between these two classes of results permits to distinguish the influence of the manipulators to move the opinion of normal agents and their ability (as stubborn group) to retain the opinion of the latter.

\begin{figure}[h]
\centering
 \includegraphics[scale=0.65]{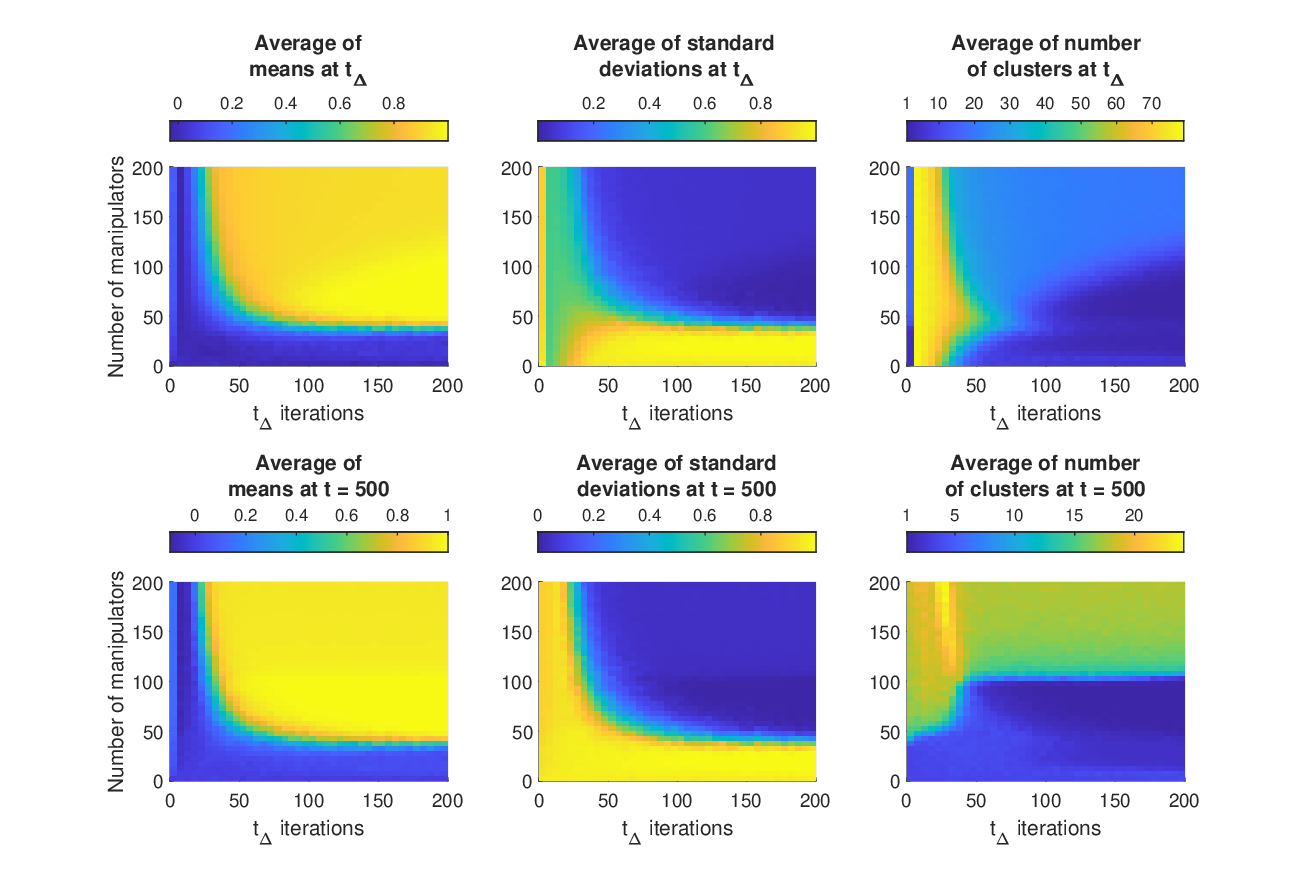}
  \caption{Simultaneous Attractive Weighted HK model: averages of mean opinions (\textit{left}), of standard deviations (\textit{center}) and of number of final clusters (\textit{right}) after 100 simulations for every $K\in \{ 5j \}_{j=0}^{40}$ and $t_{\Delta} \in \{ 5j \}_{j=1}^{40}$. The manipulative group varies its opinion from $f(0)=-0.9$ to $f(t_{\Delta})=0.9$. The confidence threshold is fixed at $\varepsilon=0.6$ constant.}
  \label{figure-scAtrNManPend1rev1_06A}
\end{figure}

\begin{figure}[h]
\centering
 \includegraphics[scale=0.65]{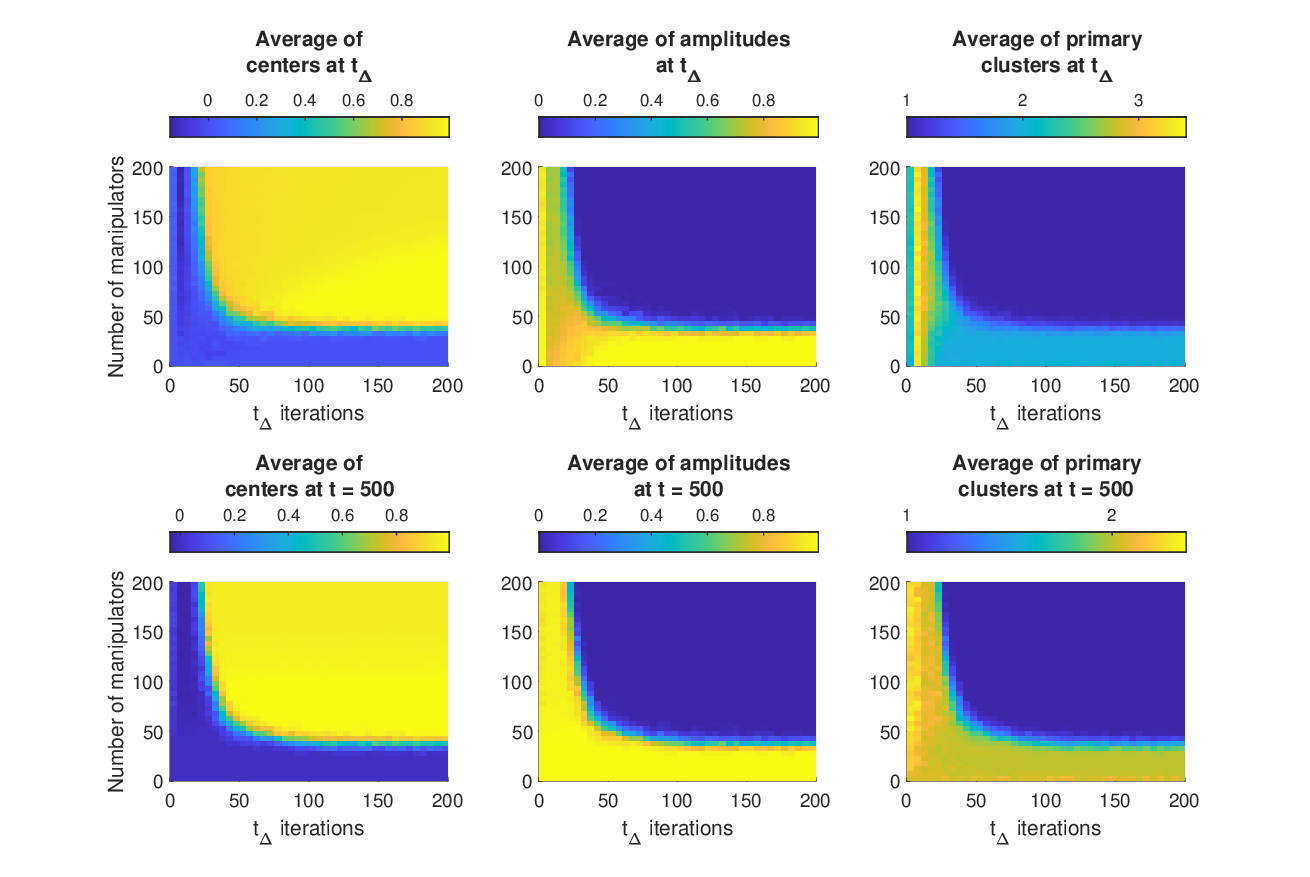}
  \caption{Simultaneous Attractive Weighted HK model: simulation with the same parameters as in figure \ref{figure-scAtrNManPend1rev1_06A}. The pictures show the average of centers (\textit{left}), of the amplitudes (\textit{center}) and of numbers of effective clusters (\textit{right}).}
  \label{figure-scAtrNManPend1rev1_06B}
\end{figure}

Notice that, in figures \ref{figure-scAtrNManPend1rev1_06A} and \ref{figure-scAtrNManPend1rev1_06B}, the upper row corresponds with the values computed at $t=t_{\Delta}$ while the lower row shows the results at $t=500$. Observe that the scale for $K\in [0,200]$ have been chosen larger than for the DW and HK--models (cf. sections \ref{sec:HKmodel} and \ref{sec:DWmodel}) due to the little influence that a small manipulative group has on the normal population. 

Taking into account the mean and the standard deviation, the influence of the manipulative group is small when $K\lesssim 50$ or $t_{\Delta} \lesssim 25$, since the averages of mean opinions are close to $x=0$ (dark blue in the left pictures). 
Their influence is important when $K\gtrsim 50$ and $t_{\Delta} \gtrsim 25$ with a narrow transition zone. 
For this values, the manipulators bring the opinion of the normal agents close to their own. Approximately if $K\in[50,100]$ with $t_{\Delta}\in [100,200]$ (bright yellow area in the left), the average opinions of normal agents are close to $x=1$ since the number of manipulative agents is not large enough to retain them in their final opinion $x=0.9$. 
If $K>100$ and $t_{\Delta}\in [100,200]$ (dark yellow in the left pictures), the manipulative group achieves that the average opinion of normal agents is around $x=0.9$, although with some dispersion $\sigma\simeq 0.2$ due to oscillating opinions. 
Moreover, when $K$ is greater than the normal population $N=100$, fragmentation of the opinion occurs.

\begin{figure}[h]
\centering
 \includegraphics[scale=0.55]{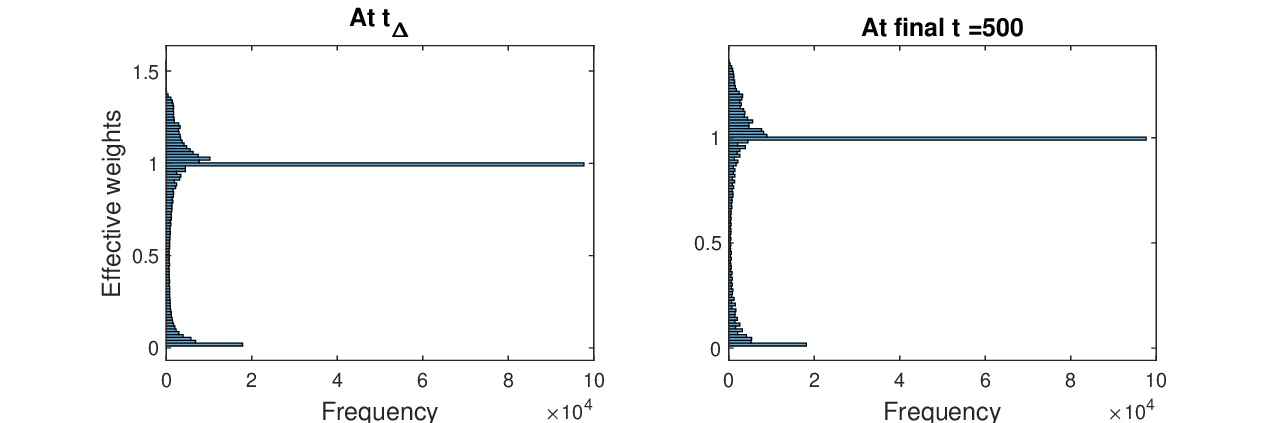}
  \caption{Simultaneous Attractive Weighted HK model: histograms of effective weights for the simulations of figures \ref{figure-scAtrNManPend1rev1_06A} and \ref{figure-scAtrNManPend1rev1_06B} at $t=t_{\Delta}$ and $t=500$.}
  \label{figure-scAtrNManPend1rev1_06C}
\end{figure}

Let us compare with the results according to the interval of primary clusters (in figure \ref{figure-scAtrNManPend1rev1_06B}). We observe that no more than 3 primary clusters reproduce the same behaviour shown for means and standard deviations.
To select the primary clusters we have chosen the filtering value $\delta=0.5$, thus eliminating the oscillating opinions that do not end up clustered. 
The histograms of effective weights are shown in figure \ref{figure-scAtrNManPend1rev1_06C}.

\bigskip

\subsubsection{Repulsive Weighted HK model with manipulators}

In the case of the Simultaneous Repulsive HK model, dynamics converge to a bimodal distribution at the extremes $x=\pm 1$ with approximately half of the normal opinions in each value. Only if the confidence threshold $\epsilon$ is small, the moderate opinions can oscillate for some iterations before convergence. 

When a manipulative group acts, normal opinions tend to separate from manipulative ones when the former are outside the latter's confidence interval. Therefore, the smaller the confidence interval, the greater the number of normal opinions that are able to drift away.

\begin{figure}[h]
\centering
 \includegraphics[scale=0.65]{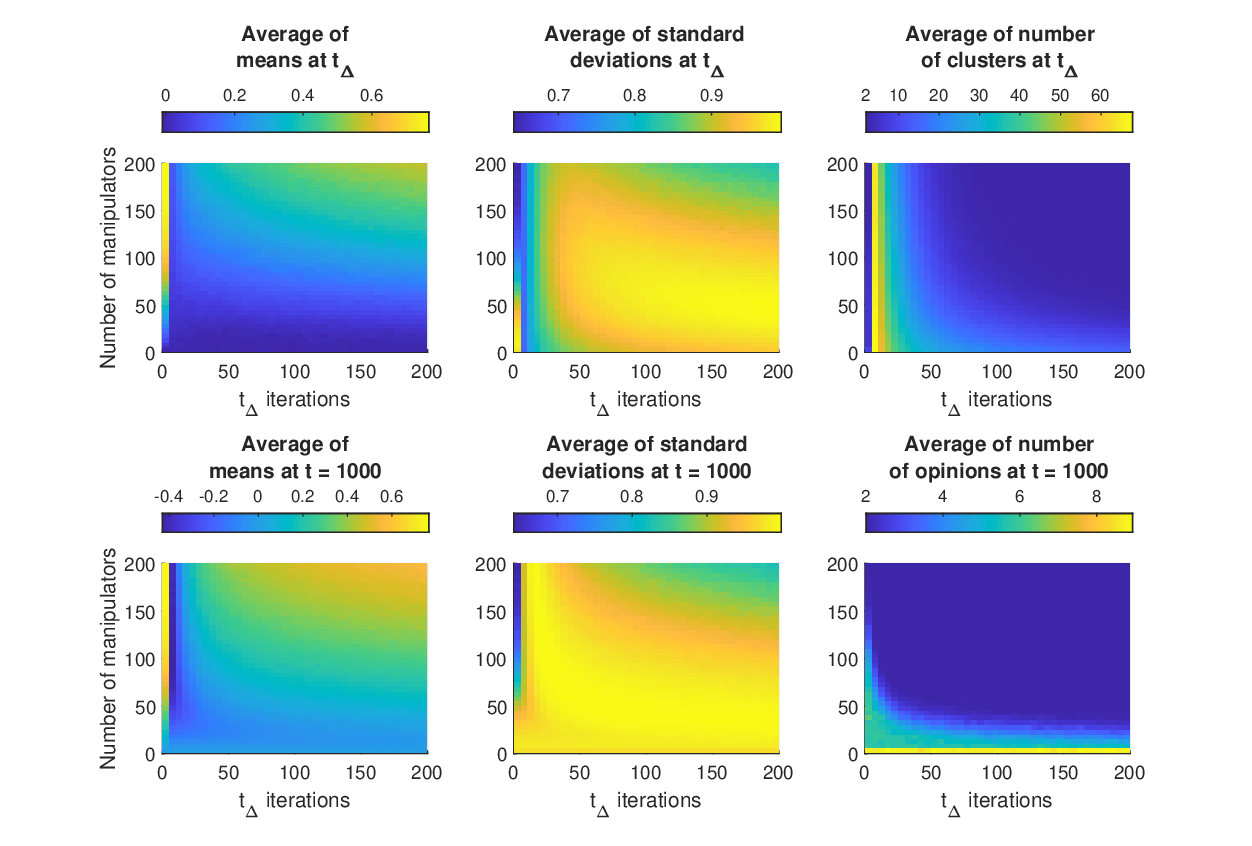}
  \caption{Simultaneous Repulsive Weighted HK model: averages of mean opinions (\textit{left}), of standard deviations (\textit{center}) and of number of final clusters (\textit{right}) after 100 simulations for every $K\in \{ 5j \}_{j=0}^{40}$ and $t_{\Delta} \in \{ 5j \}_{j=1}^{40}$. The manipulative group varies its opinion from $f(0)=-0.9$ to $f(t_{\Delta})=0.9$. The confidence threshold is fixed at $\varepsilon=0.1$ constant.}
  \label{figure-scRepNManPend2Arev1_06A}
\end{figure}

In figure \ref{figure-scRepNManPend2Arev1_06A}, we can observe that for $\varepsilon=0.1$ and $t_{\Delta}>0$, the effect of the manipulative group increases the mean values in the range of interval $[0,0.55]$ at iteration $t=t_{\Delta}$. On the other hand, at $t=1000$, when the iteration converges to the bimodal distribution $x=\pm 1$, the mean values are in $[-0.2,0.55]$ obtaining the negative values for small values of $t_{\Delta}$ and $K$. This shows that if the manipulative group changes its opinion quickly, it will get the opposite result to the one intended.

Since the final distribution converges to $x=\pm 1$, we can call $\mu$ the difference between the number of normal agents converging to $x=1$ and those converging to $x=-1$. Then we can compute the average
\[  
\overline{\mathbf{x}}= \frac{(50+\mu)-(50-\mu)}{100}=\frac{\mu}{50}
\]
then if the average satisfies $0\leq \overline{\mathbf{x}} \leq 0.55$, then $0\leq \mu \leq 28$. This implies that a manipulative group with 200 agents is not able to modify the opinion of more than 28 normal agents (28 \% of the total normal population).

In the case of $\varepsilon=0.6$, the dynamics show similar behaviour but with less ability of the manipulators to modify the opinion of the normal agents.

Notice that, since the final distribution converges to the values $x=\pm 1$ then they are primary clusters and the interval defined by those is $[-1,1]$. So, this metric do not offer further information for this model.

\subsubsection{Attractive--Repulsive Weighted HK model with manipulators}

In section \ref{sec:WHKmodel}, we have seen that the dynamics of the simultaneous Attractive--Repulsive Weighted HK model, for small values of $\varepsilon$, consists of a set of opinions oscillating in the central range of the opinion space, while extreme opinions $x=\pm 1$ can also group close normal agents together (see figure \ref{figure-scAtrRepEx2rev1}). Some central oscillating opinions can escape to converge to one of those extremal clusters.

With the intervention of manipulators, normal opinions outside the confidence interval of the manipulative group are moved away from this group. When this interval is small, the normal opinions that the manipulative group can attract are few, and this attraction can be maintained throughout the iterations when the variation of the manipulative opinions is slow, that is, when $t_{\Delta}$ is large.

\begin{figure}[h]
\centering
 \includegraphics[scale=0.6]{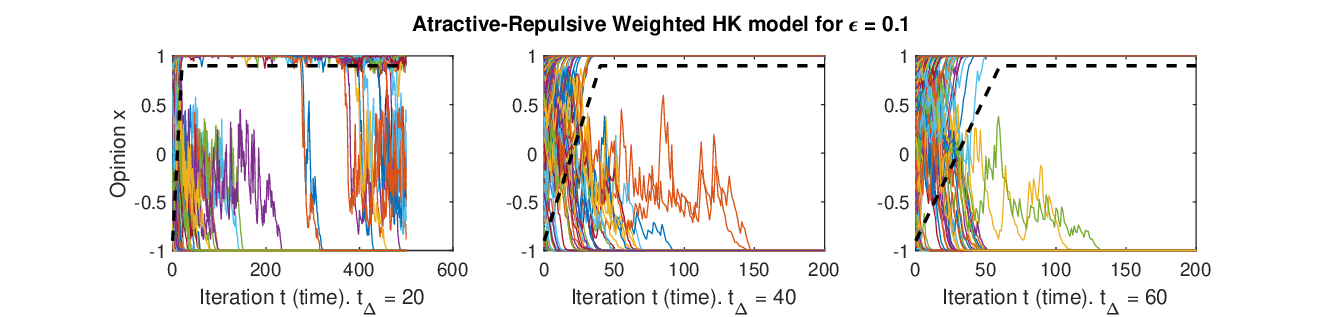}
  \caption{Simultaneous Attractive--Repulsive Weighted HK model: simulation with 100 normal agents interacting with a group of 80 manipulators varying its opinion from $f(0)=-0.9$ to $f(t_{\Delta})=0.9$ for $t_{\Delta} \in \{ 20,40,60 \}$ iterations (dashed line). The confidence threshold is fixed at $\varepsilon=0.1$ constant.}
  \label{figure-scAtrRepNMan2Arev1_01}
\end{figure}

Thus, in the simulation with 100 normal agents, a manipulative group of 80 agents whose opinions vary from $f(0)=-0.9$ to $f(t_{\Delta})=0.9$ and a confidence threshold of $\varepsilon=0.1$, it is observed (cf. figure \ref{figure-scAtrRepNMan2Arev1_01}) that, as the iterations run to $t=t_{\Delta}$, the manipulative group presses the normal opinions with higher values towards $x=1$, making their convergence to this cluster faster. 
On the other hand, the range of opinions lower than that of the manipulators becomes wider and allows normal opinions to oscillate. This oscillation will be displaced downwards by the action of the repulsive mechanism when the manipulative group becomes stubborn.
As $t_{\Delta}$ gets bigger, these oscillating opinions eventually converge to cluster $x=-1$ (cf. figure \ref{figure-scAtrRepNMan2Arev1_01}, center and right).

\begin{figure}[h]
\centering
 \includegraphics[scale=0.6]{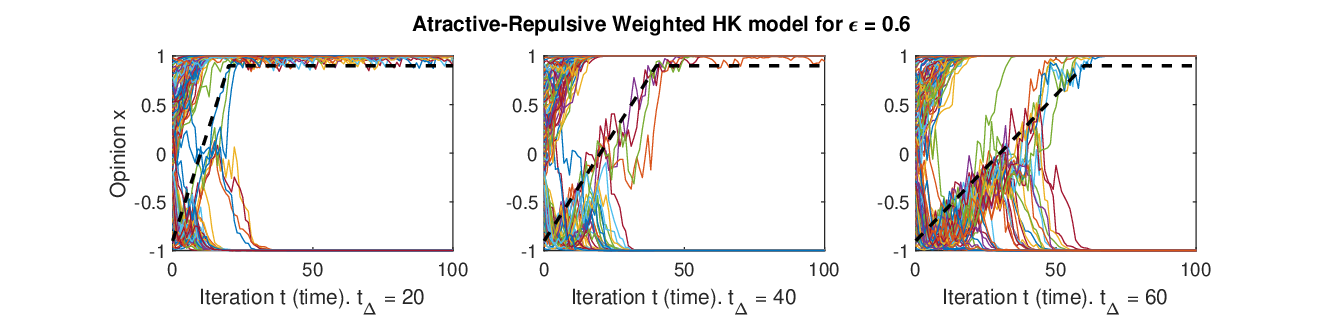}
  \caption{Simultaneous Attractive--Repulsive Weighted HK model: simulation with the same parameters as in figure \ref{figure-scAtrRepNMan2Arev1_01} except the confidence threshold, which is constant $\varepsilon=0.6$.}
  \label{figure-scAtrRepNMan2Arev1_06}
\end{figure}

When the confidence threshold $\varepsilon$ increases, the attractive mechanism also increases its influence while the repulsive decreases its own, allowing the manipulating group to attract a greater number of normal agents. This can be seen in figure \ref{figure-scAtrRepNMan2Arev1_06}, which corresponds to the simulation with the same parameters as in figure \ref{figure-scAtrRepNMan2Arev1_01} except that we have fixed $\varepsilon=0.6$.
In the three graphics of figure \ref{figure-scAtrRepNMan2Arev1_06}, we can observe that normal opinions are attracted by the manipulative group (dashed line) until the iteration $t=t_{\Delta}$ and, subsequently, when it becomes a stubborn group.

\subsubsection{Simulations for Attractive--Repulsive Weighted HK model with one manipulative group}

Now, we will consider $N=100$ equispaced normal agents in $[-1,1]$, while the variables $t_{\Delta}$ and $K$ take values in the ranges $t_{\Delta}\in\{5j\}_{j=0}^{40}$ and $K\in\{5j\}_{j=0}^{40}$.

\begin{figure}[h]
\centering
 \includegraphics[scale=0.65]{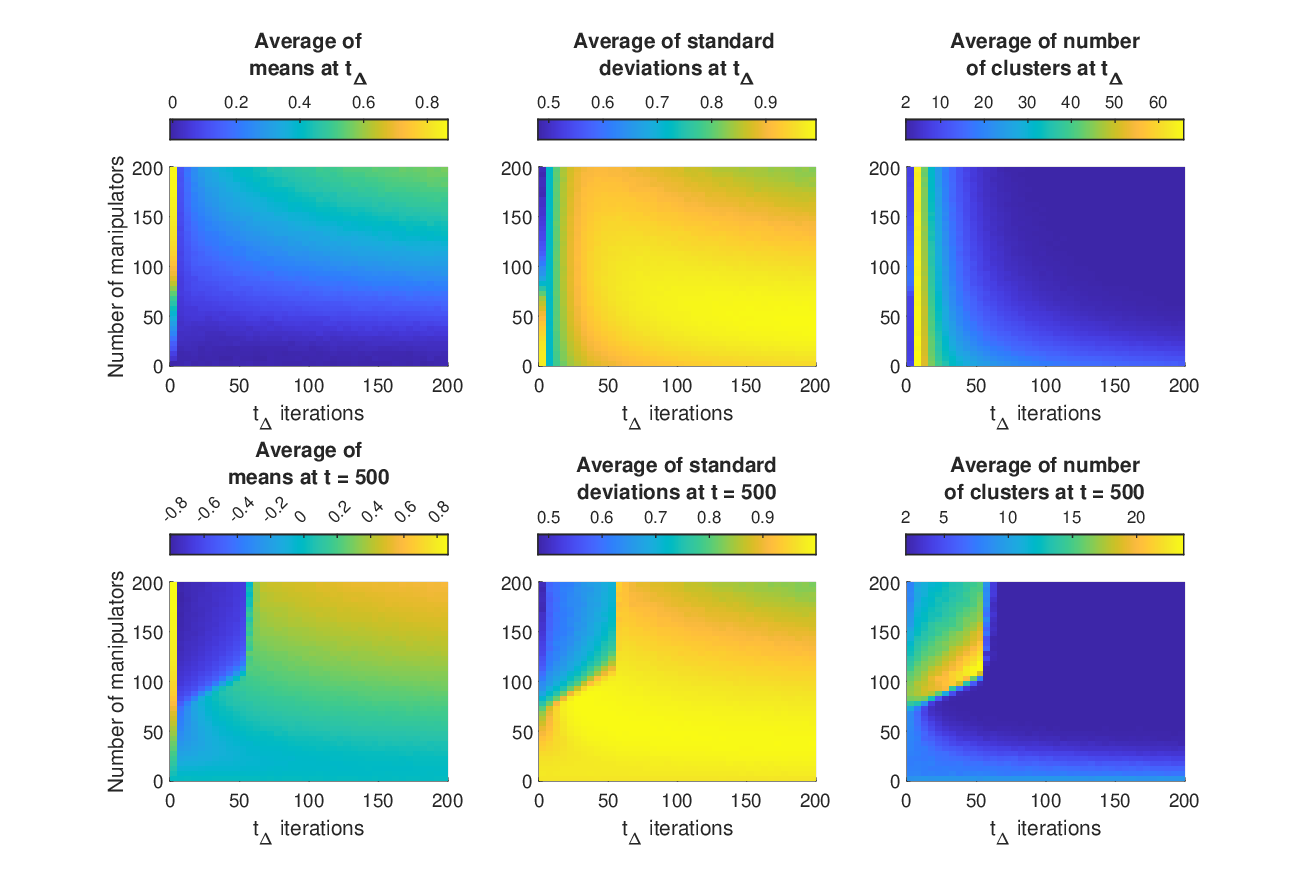}
  \caption{Simultaneous Attractive--Repulsive Weighted HK model: averages of mean opinions (\textit{left}), of standard deviations (\textit{center}) and of number of final clusters (\textit{right}) after 100 simulations for every $K\in \{ 5j \}_{j=0}^{40}$ and $t_{\Delta} \in \{ 5j \}_{j=1}^{40}$. The manipulative group varies its opinion from $f(0)=-0.9$ to $f(t_{\Delta})=0.9$. The confidence threshold is fixed at $\varepsilon=0.1$ constant.}
  \label{figure-scAtrRepNManPend2Arev1_01A}
\end{figure}

For the confidence threshold $\varepsilon=0.1$ (see figure \ref{figure-scAtrRepNManPend2Arev1_01A}), we can notice that the average of the mean opinions measured at iteration $t=t_{\Delta}$ reaches approximately the value $x\simeq 0.5$ for large values of $K$ and $t_{\Delta}$, but rarely reaches $x=0.2$ when $K \lesssim 75$.

It is important to note that the manipulative group has a significant influence on the average opinion when the size of the group is greater than 75\% of the size of the normal population, contrarily what happens in the HK and DW models, in which smaller manipulative groups can have a much more important influence.

Because the tendency to polarization in a bimodal distribution at $x=\pm 1$, even with the existence of moderate oscillating opinions, the average of the standard deviations is large and the number of clusters approaches 2 when the values of $t_{\Delta}$ cease to be small.

In the bottom row of figure \ref{figure-scAtrRepNManPend2Arev1_01A}, we show the results after $t=500$ iterations in such a way that the manipulative group becomes stubborn at $f(t)=0.9$ for all $t\in\{t_{\Delta},\ldots,500\}$. In this case, oscillating opinions (if any) tend to converge to one of the extreme clusters (or may remain oscillating), stabilizing the average values studied.

\begin{figure}[h]
\centering
 \includegraphics[scale=0.65]{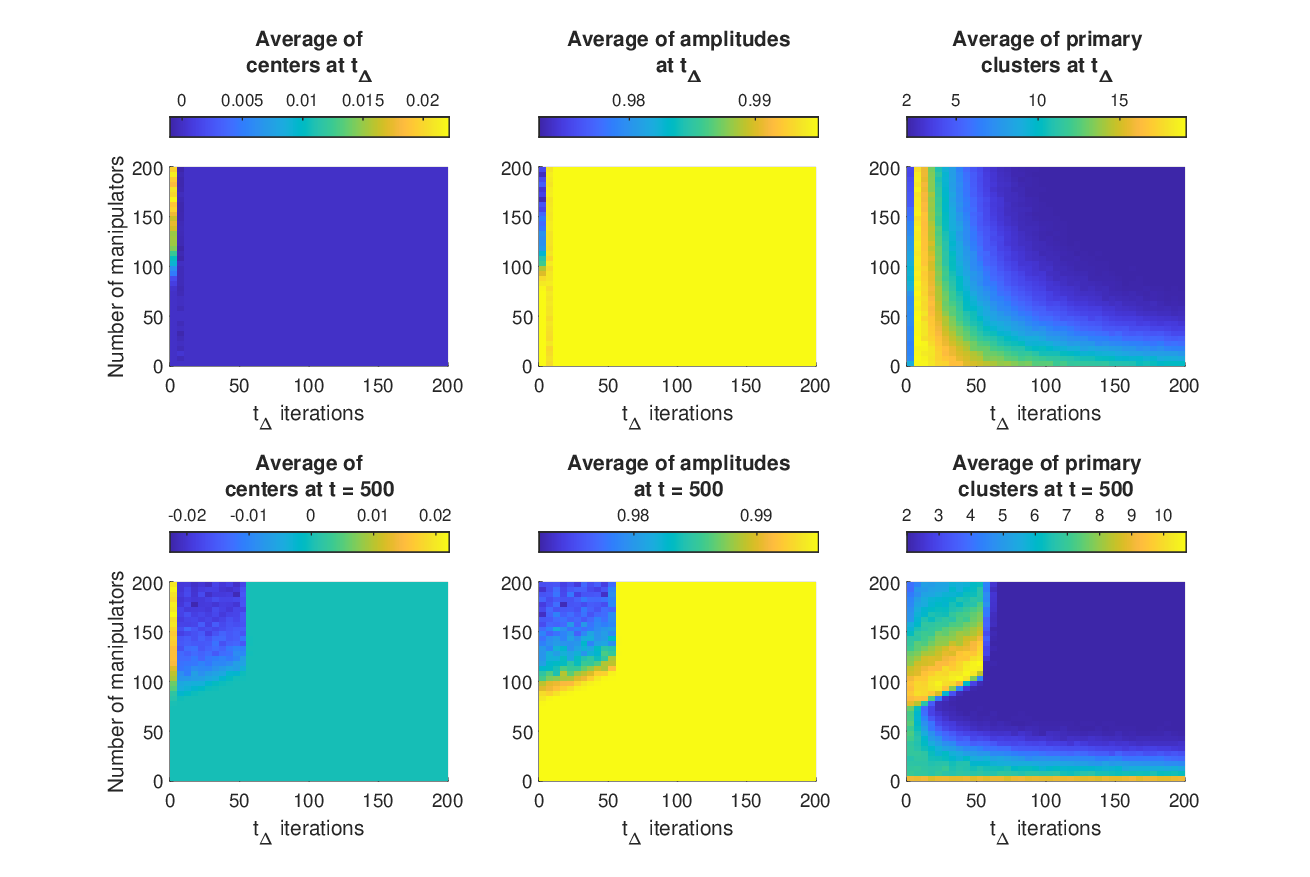}
  \caption{Simultaneous Attractive--Repulsive Weighted HK model: averages of centers, amplitudes and number of number of clusters in the interval of primary clusters with $\delta = 0$ of the same simulation of figure \ref{figure-scAtrRepNManPend2Arev1_01A}.}
  \label{figure-scAtrRepNManPend2Arev1_01B_0}
\end{figure}

Observe that when $t_{\Delta}\leq 50$ and $K\gtrsim 100$, the averages of the mean opinions move in the opposite direction to the opinion of the manipulative group, even reaching $x\simeq -0.7$ (see dark blue region in the figure \ref{figure-scAtrRepNManPend2Arev1_01A} bottom left).
Again, this is because the manipulative group changes its opinion quickly and before the normal opinions converge to one of the clusters, and since the confidence interval is small, the repulsive mechanism pushes the normal opinions in the opposite direction to that of the manipulators. 
The stabilization of the values is also observed by the reduction in the final number of clusters at $t=500$ in relation with $t=t_{\Delta}$ (see figure \ref{figure-scAtrRepNManPend2Arev1_01B_0}).

\begin{figure}[h]
\centering
 \includegraphics[scale=0.55]{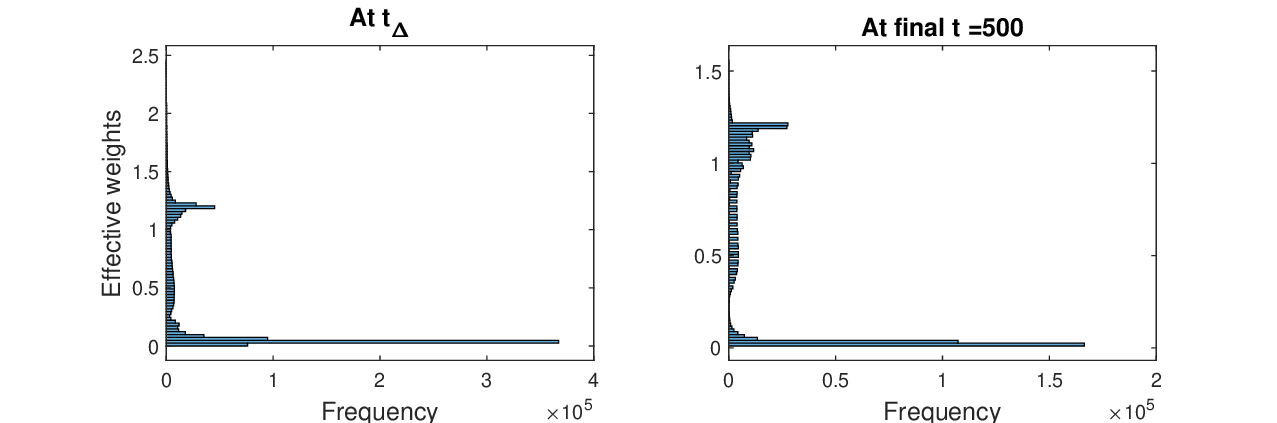}
  \caption{Simultaneous Attractive--Repulsive Weighted HK model: histograms of the effective weights in the simulation of figure \ref{figure-scAtrRepNManPend2Arev1_01A}. The most frequent effective weights are those closest to zero.  }
  \label{figure-scAtrRepNManPend2Arev1_01C}
\end{figure}

There is not an adequate cluster screening by their effective weights because the most frequent are the effective weights closest to $W\simeq 0$ (see figure \ref{figure-scAtrRepNManPend2Arev1_01C}). If we consider that each cluster is primary (filtering with the screening value $\delta =0$) then we obtain that the range of effective weights is almost the entire interval $[-1,1]$. When we choose $\delta =0.2$ for screening, we can have only one primary cluster losing the extreme cluster $x=1$ close to the opinion of the manipulators.

\begin{figure}[h]
\centering
 \includegraphics[scale=0.65]{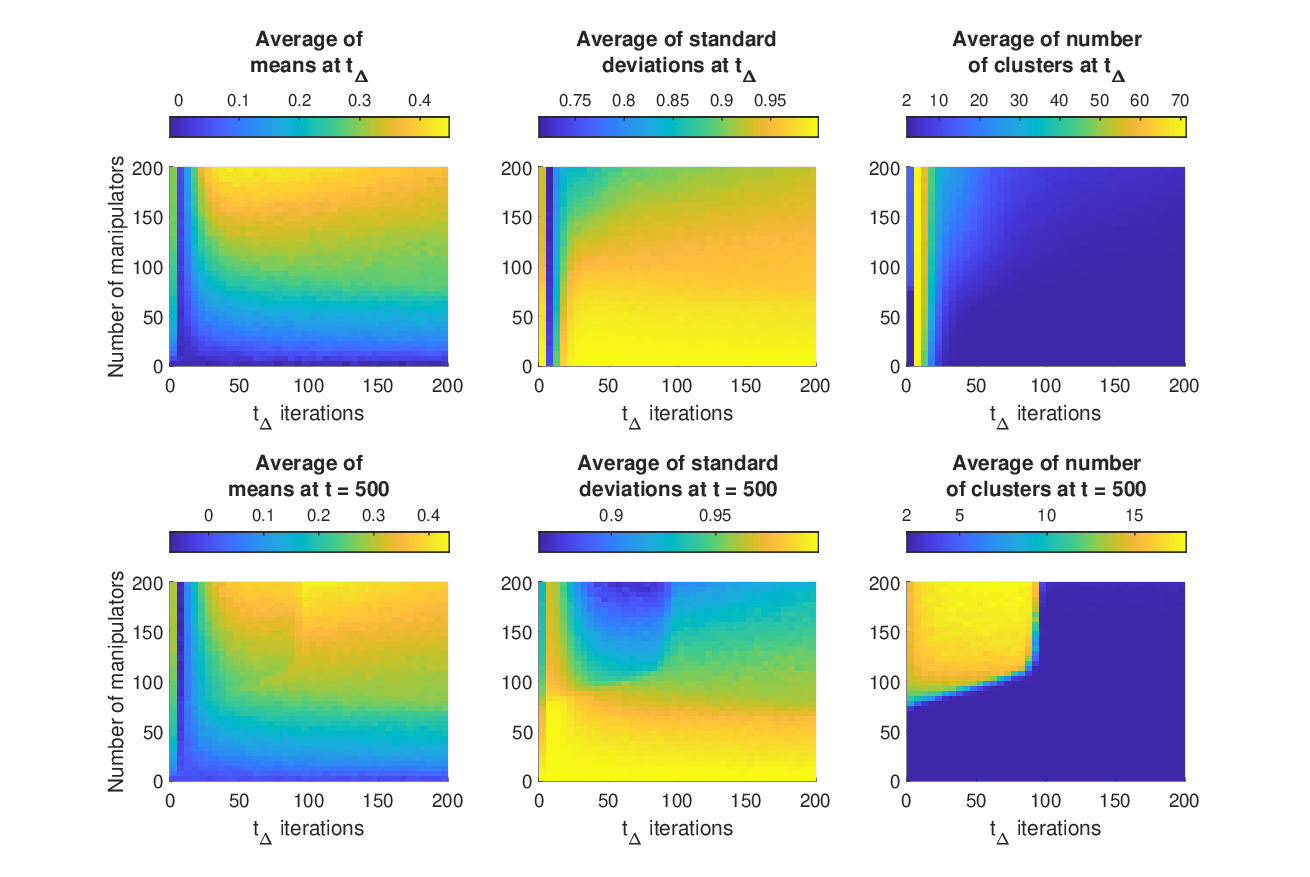}
  \caption{Simultaneous Attractive--Repulsive Weighted HK model: averages of mean opinions (\textit{left}), of standard deviations (\textit{center}) and of number of final clusters (\textit{right}) after 100 simulations for every $K\in \{ 5j \}_{j=0}^{40}$ and $t_{\Delta} \in \{ 5j \}_{j=1}^{40}$. The manipulative group varies its opinion from $f(0)=-0.9$ to $f(t_{\Delta})=0.9$. The confidence threshold is fixed at $\varepsilon=0.6$ constant.}
  \label{figure-scAtrRepNManPend2Arev1_06A}
\end{figure}

When we set the confidence threshold at $\varepsilon=0.6$, the attractive mechanism has greater influence and the manipulative group is able to attract a greater number of normal opinions. Notice that, after $t=t_{\Delta}$ iterations, the highest average of the mean opinions is reached for values $(t_{\Delta},K)\simeq (50,200)$ (in yellow in figure \ref{figure-scAtrRepNManPend2Arev1_06A} top left). 
For greater values of $t_{\Delta}$, the manipulative group needs more iterations to reach its final opinion. Thus, normal opinions in the confidence interval of the manipulative group are more likely to be repelled by other agents while they follow the manipulative group.

It is also observed, that the influence of the manipulating group is smaller that in the case $\epsilon=0.1$. In fact, the rejection effect that we found for the case $\epsilon=0.1$ does not occur, in which the opinions shifted in the opposite direction to that of the manipulating group.

\begin{figure}[h]
\centering
 \includegraphics[scale=0.55]{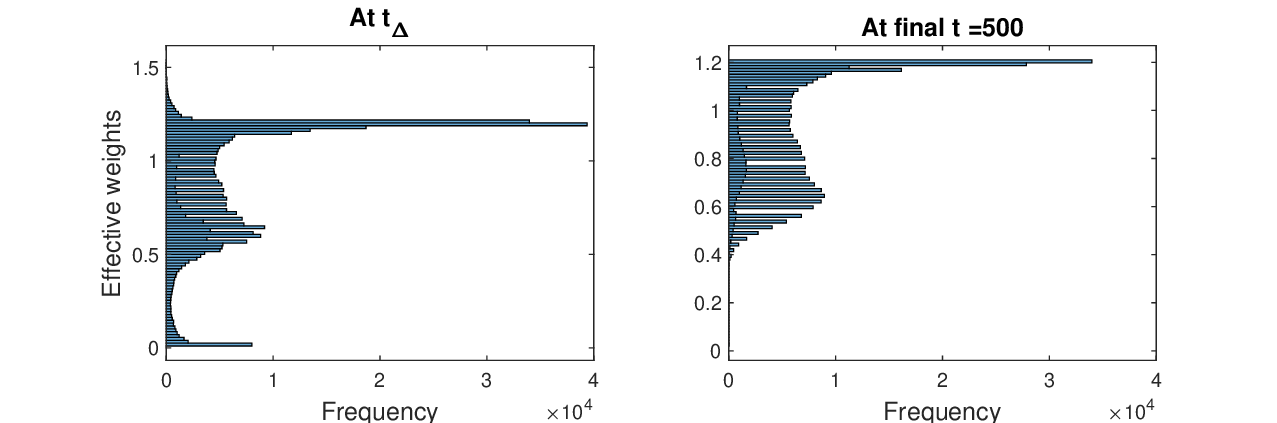}
  \caption{Simultaneous Attractive--Repulsive Weighted HK model: histograms of the effective weights in the simulation of figure \ref{figure-scAtrRepNManPend2Arev1_06A}.}
  \label{figure-scAtrRepNManPend2Arev1_06C}
\end{figure}

Now, it is possible to screen by primary clusters since the most frequent effective weights are close to $W=1.2$ (see histogram in figure \ref{figure-scAtrRepNManPend2Arev1_06C}). 

\begin{figure}[h]
\centering
 \includegraphics[scale=0.65]{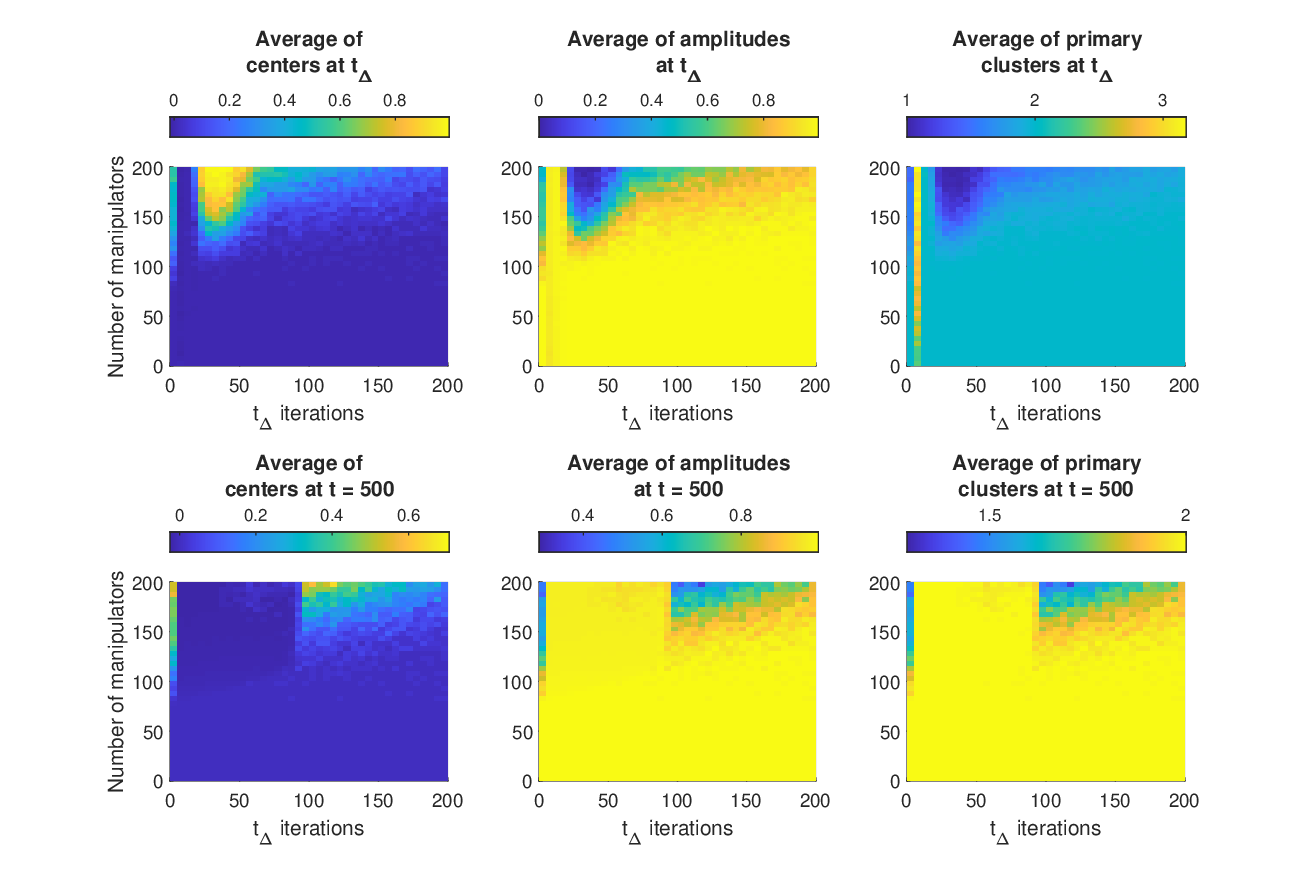}
  \caption{Simultaneous Attractive--Repulsive Weighted HK model: average of centers, amplitudes and number of number of clusters in the interval of primary clusters with $\delta = 0.5$ of the same simulation of figure \ref{figure-scAtrRepNManPend2Arev1_06A}.}
  \label{figure-scAtrRepNManPend2Arev1_06B}
\end{figure}

If we choose the screening value $\delta=0.5$ at the iteration $t=t_{\Delta}$, we observe that for $K\leq 100$, we get two primary clusters at $x=\pm1$ and only for values of $K>150$ and $t_{\Delta}$ small, we obtain an only primary cluster (yellow region at top left of figure \ref{figure-scAtrRepNManPend2Arev1_06B}). But, in this region, the final distribution at $t=500$ iterations converges to two primary clusters.

\section{Conclusions and discussion}\label{sec:conclusion}

We have studied the opinion dynamics in various models of bounded confidence under the influence of one group of manipulative agents.
This type of agent, as a generalization or extension of stubborn agents, allows them to have influence within the union of the confidence intervals corresponding to their range of opinions.
Thus, strategies can be set up to modify the natural dynamics of the model (due only to normal agents). 
This modification can cause socially acceptable opinions of normal agents to become extreme or vice versa. The range of opinions can shift as well as expand or contract in the opinion space, reminiscent of Overton window theory.
Therefore, manipulators can try to modify the Overton window to form a new framework of acceptable opinions.

In this article, we have described the strategies of one single manipulative group that linearly varies its opinions over a wide range of opinions, and then becomes a stubborn group until convergence or stabilization of the dynamics.
The number of manipulators in the group, as well as the rate of variation of their opinions, have an unequal influence on the different models studied, with the different confidence thresholds.
Due to the arbitrariness of choosing a definition for the Overton window, we consider that the mean as the central point and the standard deviation as the width can be used to determine it. 
In some cases, when possible, we have used the center and amplitude of the interval of primary clusters to measure the influence of manipulators, with similar results. 
We have shown examples of simulations of opinion dynamics with a group of manipulators interpreting and quantifying their influence on normal agents as a function of the variables defining that group. 
In the HK and DW models, due to their tendency to central clustering, a manipulative group with a sufficient number of agents and/or whose change of opinion is sufficiently slow, can attract a large number of normal agents and take them with them anywhere in the opinion space.
Moreover, for the simultaneous HK--model, the conditions of these variables have been fully determined so that manipulators attract the opinions of another group of normal agents.
In Weighted HK models, a larger number of manipulators is needed to influence the opinion of normal agents, and rapid changes in the opinion of manipulators could have the opposite effect on normal agents.

%\bigskip

There are still pending issues that will be studied in future articles, among them, the competition between various groups of manipulators according to their size and the trajectories of their opinions, the finding of new models that adapt to the movement of the Overton window placing it in the central point of the space of opinions. In this sense, the density function $\phi_i(t)=1-\vert x_i(t) \vert$ in the iteration of the Weighted HK model can be modified so that its mean in each iteration matches the central point (mean or center) of the distribution of normal opinions.

\bigskip

All the simulations and graphs in this article have been made using the R2020b and R2024a versions of Matlab.

\section*{Acknowledgements}

The author would like to thank the two anonymous referees for their comments, especially on the structuring of the article and for the incorporation of the primary clusters metric, and also Gemma Rioja for the suggestions provided on the wording. These contributions have significantly improved this article.

\section*{Data availability}

No new data were created or analysed during this study. Data sharing is not applicable to this article.

%%%%%%%%%%%%%%%%%%%%%%%%%%%%%%%%%%%%%%%%%%%%%%%%%%%%%

\end{document}